\documentclass[a4paper,11pt, twoside]{article}
\usepackage{url}
\usepackage{amsmath}
\usepackage{booktabs}
\usepackage{color}
\usepackage{graphicx}
\usepackage{todonotes}
\usepackage{mathtools}
\usepackage{subcaption}
\usepackage{lipsum}
\usepackage{epstopdf}
\usepackage{hyperref}
\usepackage[left=2.5cm, right=2.5cm, top=2.3cm]{geometry}
\usepackage{amssymb}
\usepackage{amsthm}
\usepackage{mathtools}
\usepackage{listings}
\usepackage{epsfig}
\usepackage{latexsym}
\usepackage{mathtools}
\usepackage{booktabs}
\usepackage{amsfonts}
\usepackage{hyperref}
\usepackage[mathscr]{euscript}
\usepackage{bm}
\usepackage{paralist}
\usepackage{enumitem}
\usepackage[backend=bibtex,style=numeric,citestyle=numeric,giveninits=true,sortcites=true,maxbibnames=10]{biblatex}

\usepackage{filecontents}

\immediate\write18{bibtex \jobname}
\definecolor{deepblue}{rgb}{0,0,0.5}
\definecolor{deepred}{rgb}{0.6,0,0}
\definecolor{deepgreen}{rgb}{0,0.5,0}

\definecolor{DarkBlue}{rgb}{0.00,0.00,0.55}
\definecolor{Black}{rgb}{0.00,0.00,0.00}
\hypersetup{
	linkcolor = DarkBlue,
	anchorcolor = DarkBlue,
	citecolor = DarkBlue,
	filecolor = DarkBlue,
	colorlinks  = true,
	urlcolor = Black
}

\graphicspath{ {./Figures/},{./Figures/Brain_MCQMC/},{./} }
\DeclareGraphicsExtensions{.eps,.pdf,.png,.jpg} 

\usepackage{todonotes}
\definecolor{myblue}{RGB}{135, 206, 250}
\definecolor{mygreen}{RGB}{100, 220, 0}

\DeclareMathOperator{\Div}{div}

\DeclareMathOperator{\W}{\dot{W}}
\newcommand{\inner}[2]{( #1, #2 )}

\newcommand{\R}{\mathbb{R}}

\newcommand{\dx}{\, \mathrm{d} \bm{x}}

\newcommand{\E}{\mathbb{E}}
\newcommand{\V}{\mathbb{V}}

\newtheorem{theorem}{Theorem}[section]
\newtheorem{remark}{Remark}[section]

\begin{document}
\title{Fast uncertainty quantification of tracer distribution in the brain interstitial fluid with multilevel and quasi Monte Carlo}

\providecommand{\keywords}[1]{\textbf{Key words:} #1}

\author{
	M.~Croci\thanks{Mathematical Institute, University of Oxford, Oxford, UK. Department for Numerical Analysis and Scientific Computing, Simula Research Laboratory, Lysaker, Norway (\textbf{\url{matteo.croci@maths.ox.ac.uk}}).}
	\and
	V.~Vinje\thanks{Department for Numerical Analysis and Scientific Computing, Simula Research Laboratory, Lysaker, Norway (\textbf{\url{vegard@simula.no}}).}
	\and
	M.~E.~Rognes\thanks{Department for Numerical Analysis and Scientific Computing, Simula Research Laboratory, Lysaker, Norway (\textbf{\url{meg@simula.no}}).}
}

\maketitle


\begin{abstract}
	Efficient uncertainty quantification algorithms are key to
	understand the propagation of uncertainty -- from uncertain input
	parameters to uncertain output quantities -- in high resolution
	mathematical models of brain physiology. Advanced Monte Carlo
	methods such as quasi Monte Carlo (QMC) and multilevel Monte Carlo
	(MLMC) have the potential to dramatically improve upon standard
	Monte Carlo (MC) methods, but their applicability and performance in
	biomedical applications is underexplored. In this paper, we design
	and apply QMC and MLMC methods to quantify uncertainty in a
	convection-diffusion model of tracer transport within the brain. We
	show that QMC outperforms standard MC simulations when the number of
	random inputs is small. MLMC considerably outperforms both QMC and
	standard MC methods and should therefore be preferred for brain
	transport models.
\end{abstract}


\maketitle


\section{Introduction}

Mathematical models in biology involve many parameters that are uncertain or in some cases unknown. Over the last years, increased computing power has expanded the complexity and increased the number of degrees of freedom of many such models. For instance, full scale modeling of the brain requires very high spatial resolution to resolve the details on the curvature of the surface. For this reason, efficient uncertainty quantification algorithms are now needed to explore the often large parameter space of a given model. Advanced Monte Carlo methods such as quasi Monte Carlo (QMC) and multilevel Monte Carlo (MLMC) have become very popular in the mathematical, engineering, and financial literature for the quantification of uncertainty in model predictions. However, applying these methods to physiologically relevant brain simulations is a difficult task given the typical complexity of the models and geometries involved. 
  
Efficient elimination of waste products from the brain is crucial for a functional central nervous system. In humans, the brain is responsible for around 20\% of the total oxygen consumption and 15\% of the cardiac output~\cite{sokoloff1989circulation}. Despite this high energy demand, the brain lacks lymphatic vessels, which carry waste products in the rest of the body, thus introducing the need for an effective alternative mechanism. Proposed theories of waste clearance all include a combination of diffusion and convection to clear substances out of the brain. There is also consensus that interstitial fluid (ISF) within the brain can exchange with cerebrospinal fluid (CSF) surrounding the brain to transfer waste products. However, there is disagreement on where the CSF/ISF-exchange occurs, in which direction it occurs, the forces involved in transport and how much of the transport within the brain that can be explained by diffusion alone~\cite{IliffEtAl2012,JessenEtAl2015,IliffEtAl2013,CarareEtAl2008,aldea2019cerebrovascular,SmithEtAl2017}.

Interchange between CSF and ISF is also an important mechanism for effective drug delivery to the brain by substance administration in the SAS. The blood-brain-barrier is one of the main obstacles for this type of drug delivery to the brain~\cite{wolak2013diffusion}. The contrast agent gadobutrol has been used as a CSF tracer to assess transport into brain tissue, and possibly across the blood-brain-barrier~\cite{RingstadEtAl2017, RingstadEtAl2018}. However, human studies are sparse, the number of subjects are limited, and MRI can not directly capture phenomena on the microscale (e.g. transport across the blood-brain-barrier). 
 
Mathematical modeling is therefore a compelling tool to investigate CSF/IFS-exchange related to brain clearance or drug delivery. Computational studies~\cite{AsgariEtAl2016, sharp2019dispersion, daversin2020mechanisms} investigating CSF flow in paravascular spaces (PVS) have not been able to reproduce the average flow velocities reported in experimental studies with microspheres~\cite{mestre2018flow, BedussiEtAl2017}. Within the interstitium, both experimental~\cite{SmithEtAl2017}, and computational~\cite{HolterEtAl2017} studies have concluded that diffusion dominates convection. Still, tracer experiments in human beings~\cite{RingstadEtAl2017, RingstadEtAl2018} show transport of gadobutrol to the brain that is unlikely to be explained by diffusion alone~\cite{CrociVinjeRognes2019}. 

The aforementioned mathematical models involve parameters that are usually unknown to some degree. Therefore, evaluating the models' parameter sensitivities is vital to give confidence in the model predictions and conclusions. For instance, brain tissue permeability range over several orders of magnitude in the literature~\cite{HolterEtAl2017, smith2007interstitial}. Recently, Fultz et al.~\cite{fultz2019coupled} found coherent oscillations of electrophysiological waves, blood flow and cerebrospinal fluid flow during non-rapid eye movement sleep, suggesting that these processes are all interlinked. Full-scale modeling of CSF/ISF flow in the brain thus introduces several more interlinked parameters related to these additional processes. Previous studies in the literature that have included extensive parameter sensitivity analyses have been computationally cheap \cite{Bilston2003,asgari2015astrocyte}, allowing for  parameter space exploration within a reasonable amount of time. However, when the model of interest is expensive to simulate, more advanced uncertainty quantification (UQ) methods are needed.

Standard Monte Carlo (MC) methods have successfully been applied to simulations of e.g.~the cardiovascular system~\cite{BiehlerEtAl2015,QuaglinoEtAl2018,Quaglino2018}, and to brain solute transport~\cite{CrociVinjeRognes2019}. However, when working with physiologically realistic, MRI-derived geometries, standard MC methods are typically prohibitively costly and more advanced methods, e.g.~quasi Monte Carlo (QMC) or multilevel methods (multilevel, multilevel quasi, or multi-fidelity Monte Carlo), are to be preferred. We remark that each of these methods bring their own additional requirements: QMC requires either the output functionals of interest to have low effective dimensionality with respect to the random input and/or the input dimensions to be ordered in order of decaying importance~\cite{CaflishMorokoffOwen}. Whether this assumption is satisfied is strongly problem-dependent. On the other hand for Multilevel Monte Carlo (MLMC) to work well, a good statistical coupling must be enforced between the levels and a mesh hierarchy on which a good rate of bias and variance decay can be appreciated is needed~\cite{giles2015multilevel}. Even though such a hierarchy can always be obtained by refining and/or coarsening a given mesh in theory, this is far from trivial in practice. A multi-fidelity Monte Carlo (MFMC)~\cite{peherstorfer2018survey} approach can be beneficial in this case, since it can incorporate in the hierarchy low-fidelity models that are still correlated with the fine-mesh simulations, but do not require a computational grid. However, suitable low-fidelity models are not always available in practice. Finally, multilevel quasi Monte Carlo (MLQMC) methods combine the advantages and the requirements of both QMC and MLMC and, as such, may be advantageous only in the cases in which both QMC and MLMC perform well.

In light of the above considerations, the efficiency, feasibility and
performance of enhanced Monte Carlo methods such as QMC, MLMC, MFMC,
and MLQMC is highly problem-dependent. Moreover, determining a-priori
which method that will bring the largest performance benefits for a
given problem is highly nontrivial. As the model complexity and
dimension grows, choosing the most efficient UQ method becomes even
more important. Additionally, knowing when QMC and MLMC both work well
provides a model domain for which MLQMC methods can bring substantial
additional computational improvements. Generally speaking, correctly
setting up a QMC or MLMC method in complex geometries is a difficult
task, and to our knowledge QMC and MLMC methods have not yet been
employed for UQ in simulations of brain physiology.

In this study, we therefore investigate different enhanced Monte Carlo approaches for quantifying uncertainty in a model with random field coefficients deriving from MRI-studies in humans~\cite{RingstadEtAl2017,RingstadEtAl2018}. The model, and its connection to tracer experiments in humans, has previously been described in detail~\cite{CrociVinjeRognes2019}, as well as the clinical relevance of the simulated tracer experiments in~\cite{RingstadEtAl2017}. We introduce and evaluate the relative performance of different Monte Carlo methods, namely standard, quasi and multilevel Monte Carlo methods~\cite{Croci2018,CrociMLQMC} as implemented in the FEMLMC library. The software platform used is comprised of generally available open-source software libraries (FEniCS~\cite{AlnaesBlechta2015a}, libsupermesh~\cite{libsupermesh,libsupermesh-tech-report}, and FEMLMC~\cite{femlmc}).

Our findings show that QMC outperforms standard Monte Carlo, but only when the number of (in this case infinite-dimensional) random inputs is small, bringing a 10-fold improvement in the computational cost. On the other hand, MLMC always outperforms both standard Monte Carlo and QMC, and is two orders of magnitude faster than standard Monte Carlo, showing that MLMC should be the method of choice when solving brain transport problems. Our results further suggest that problems with a small number of random field inputs might be amenable to a MLQMC treatment.

This paper is organized as follows. In Section \ref{sec:background}, we give a brief overview of mathematical, numerical and algorithmic background. In Section \ref{sec:methods}, we present the baseline stochastic model for brain tracer movement, as originally introduced in~\cite{CrociVinjeRognes2019}, and in Section \ref{sec:coefficient_models} we formalize two UQ test problems building on this baseline model. We detail the numerical and computational solution algorithms in Section \ref{sec:numerical_details_stochastic_models} before presenting numerical results in Section \ref{sec:num_brain_results}. We discuss our findings in Section~\ref{sec:discussion} and conclude in Section~\ref{sec:conclusions}.

\section{Monte Carlo methods and stochastic sampling}
\label{sec:background}

\subsection{Preliminaries}
\label{sec:preliminaries}

In what follows we indicate by $\bm{x} \in G \subset \R^3$ a given spatial point in a domain $G$, with $t>0$ a time point, and with $\omega$ a given probabilistic event, living in a sample space $\Omega$. For example, we might indicate with $f(t,\bm{x})$ a generic function of time and space and with $z(\omega)$ a generic random variable of expected value $\E[z]$ and variance $\V[z]$. Additionally, we will consider random functions of space, varying both in space and across random realisations. These are called random fields and we use the notation e.g.~$u(\bm{x},\omega)$. More formally, a random field $u(\bm{x},\omega)$ is a collection of random variables such that each point value $u(\bm{x},\cdot)$ is a random variable for every $\bm{x}$ and $u(\cdot,\omega)$ is a function of space for fixed $\omega$.

For Gaussian fields, all point values (which are random variables) are jointly Gaussian, and a Gaussian field $u(\bm{x},\omega)$ is uniquely determined by prescribing a mean function $\mu(\bm{x})=\E[u(\bm{x},\omega)]$ and a covariance function $C(\bm{x},\bm{y})=\E[(u(\bm{x},\omega)-\mu(\bm{x}))(u(\bm{y},\omega)-\mu(\bm{y}))]$. A ubiquitous family of Gaussian fields is the Mat\'ern family: a Mat\'ern field is a Gaussian field with covariance of the type
\begin{align}
\label{eq:Matern}
C(\bm{x},\bm{y}) = \dfrac{\sigma^2}{2^{\nu-1}\Gamma(\nu)}(\kappa r)^\nu \mathcal{K}_\nu(\kappa r),\ \ r=\Vert \bm{x}-\bm{y}\Vert _2,\ \ \kappa = \frac{\sqrt{8\nu}}{\lambda},\ \ \bm{x},\bm{y}\in \R^d,
\end{align}
where $\sigma^2$, $\nu$, $\lambda > 0$ is the variance, smoothness parameter and correlation length of the field, respectively. Here $\Gamma(x)$ is the Euler Gamma function and $\mathcal{K}_\nu(x)$ is the modified Bessel function of the second kind. The $\sigma^2$, $\nu$ and $\lambda$ parameters determine the characteristics of the random field. For instance, in our model the random field represents CSF/ISF-flow and the correlation length was chosen to match approximately the distance between larger arteries and veins, to be able to model CSF inflow along arteries and outflow along veins~\cite{IliffEtAl2012}.

In this paper we consider the solution of a convection-diffusion-reaction equation with random fields as coefficients for the movement of tracer within the brain. The convection-diffusion-reaction equation reads as: find the tracer concentration $c = c(t, \bm{x}, \omega)$
for $\bm{x} \in G$, $\omega \in \Omega$ and $t \geq 0$ such that
\begin{equation}
\label{eq:generic}
\dot c(t,\bm{x}, \omega)
+ \nabla \cdot (\bm{v}(\bm{x},\omega) c(t,\bm{x},\omega))
- \nabla \cdot (D^{\ast}(\bm{x},\omega) \nabla c(t,\bm{x},\omega)) + rc(t,\bm{x},\omega)= 0.
\end{equation}
Here, the domain $G\subset\R^3$ represents the brain, the superimposed dot represents the time derivative, $D^{\ast}$ is the effective diffusion coefficient of the tracer in the tissue, $\bm{v}$ represents a convective fluid
velocity and $r \geq 0$ is a drainage coefficient. Further details, including boundary and initial conditions, will be presented in Section~\ref{sec:methods}.

The solution $c$ is random and varies in time and space as well, while the coefficients $\bm{v}(\bm{x},\omega)$ and $D^{\ast}(\bm{x},\omega)$ are random fields. A typical assumption is that one is interested in computing one (or more) output functional of interest $Q(\omega) = \mathscr{Q}(c)$, e.g.~$\mathscr{Q}$ could be the spatial average of $c$ over a region of interest. Typically, solving \eqref{eq:generic} then means to compute the expected value of $Q(\omega)$. Note that most other statistics (such as the variance or the cumulative density function (CDF)) can be rewritten as an expectation and thus computed analogously.

A basic method for solving \eqref{eq:generic} is the standard Monte Carlo (MC) method, for which the expectation is approximated by the sample average
\begin{align}
\E[Q]\approx\hat{Q}=\frac{1}{N}\sum\limits_{i=1}^NQ(\omega_i),
\end{align}
obtained with $0<N\in\mathbb{N}$ samples of $Q(\omega)$. Note that each sample of $Q$ requires computing samples of the coefficients $\bm{v}(\bm{x},\omega)$, $D^{\ast}(\bm{x},\omega)$ and the corresponding solution of \eqref{eq:generic}. Alas, the MC method converges slowly in terms of the number of samples $N$, namely at a rate of $\mathcal{O}(N^{-1/2})$. This makes standard MC quite expensive: assuming an $\varepsilon^2$ tolerance for the mean square error (MSE) $\E[(Q-\hat{Q})^2]$ and a cost per sample of $\varepsilon^{-q}$ for some positive problem-specific number $q$, standard MC has a total cost complexity of $O(\varepsilon^{-2-q})$ \cite{giles2015multilevel}. In practice, more advanced methods, such as Quasi Monte Carlo (QMC), multilevel Monte Carlo (MLMC) and multilevel quasi Monte Carlo (MLQMC), are to be preferred. In what follows we give a brief description of the QMC and MLMC methods.

\subsection{Quasi Monte Carlo}
We now give a quick overview of the quasi Monte Carlo method and we refer the reader to the book by Lemieux \cite{Lemieux2009} for a more in-depth description. The advantage of QMC with respect to standard MC is that in QMC methods, the convergence rate with respect to the number of samples $N$ is improved from $O(N^{-1/2})$ to $O(N^{-1+\epsilon})$ for any $\epsilon>0$. At the heart of QMC is the approximation or reinterpretation of the expectation as a high-dimensional integral over the unit hypercube:
\begin{equation}
  \E[Q]= \int_\Omega Q(\omega) \, \text{d}\mathbb{P}(\omega) \approx \int_{[0,1]^s} H(\bm{x}) \dx,
\end{equation}
where $\mathbb{P}$ is a suitable probability measure, $H(\bm{x})$ for $\bm{x}\in\R^s$ is some suitable function and $s$ is typically the dimensionality of the random input needed to sample $Q$. The QMC method can then be expressed as a quadrature rule over $[0,1]^s$ with equal weights, approximating the right-hand side integral with
\begin{align}
I = \int_{[0,1]^s} H(\bm{x}) \dx \approx \frac{1}{N} \sum_{n=1}^N H(\bm{x}_n) = I_N,
\end{align}
with $\bm{x}_n\in\R^s$. Choosing the $\{\bm{x}_n\}_{n=1}^N$ uniformly at random results in a convergence rate of $O(N^{-1/2})$. However, there exist deterministic point sequences, so-called low-discrepancy point sequences, that can yield an improved rate of $O(N^{-1}(\log N)^s)$ \cite{Morokoff1995}. This is the key idea of QMC methods: estimating the integral $I$ by a low-discrepancy sequence.

Now, let $\V[Q]$ be the variance of $Q$. In standard MC, the statistical error is
$O(\V[Q]/N)^{1/2}$ (owing to the central limit theorem) and $\V[Q]$ can be estimated by taking the sample variance of $Q$. However, in standard QMC, the point sequence $\{\bm{x}_n\}_{n=1}^N$ is deterministic and therefore there is no notion of statistical error available. For this reason, a practical estimate for the approximation error introduced by QMC is not available. A common solution is to randomize the sequence (while still preserving the hypercube-covering properties) in order to retain a measure of estimator variability (see e.g.~Chapter 6 of \cite{Lemieux2009} for an overview and \cite{Owen2003} for a comparison of different randomization strategies). This yields a randomized QMC method: let $\{\hat{\bm{x}}_{n,m}(\omega)\}_{n=1,m=1}^{n=N,m=M}$ be $M$ independent randomizations of a low-discrepancy sequence $\{\bm{x}_n\}_{n=1}^N$, then randomized QMC estimates $I$ as
\begin{align}
\label{eq:randomized_QMC}
I \approx \hat{I}_{M,N}(\omega) = \frac{1}{M}\sum\limits_{m=1}^M I^m_N(\omega) = \frac{1}{M}\sum\limits_{m=1}^M \left(\frac{1}{N}\sum\limits_{n=1}^NH(\hat{\bm{x}}_{n,m}(\omega))\right),
\end{align}
where each of the $I_N^m(\omega)$ are now random and a confidence interval can therefore be estimated provided $M$ is large enough. In this paper we use $M=32$. Assuming a fixed $M$, a given $MSE$ tolerance $\varepsilon^2$, a $O(\varepsilon^{-q})$ cost per sample and a QMC convergence order of $O(N^{-1+\epsilon})$ for any $\epsilon>0$, the total cost of randomized QMC is $O(\varepsilon^{-q-1/(1-\epsilon)})$. If we take $\epsilon$ to be small, this is almost $\varepsilon^{-1}$ times better than standard MC.

\begin{remark}
  \label{rem:low_effective_dimensionality}
  The $(\log N)^s$ can dominate the early convergence behaviour of QMC methods, in which case the suboptimal $O(N^{-1/2})$ convergence is observed initially. This behaviour is typically exacerbated when the random input dimensionality $s$ is particularly large (e.g.~in applications with infinite-dimensional random inputs such as random fields), to the extent that the faster rate is never observed in practice. However, this is not always the case: if the QMC integrand $H(\bm{x})$ is inherently lower dimensional in the sense that it can be well approximated by a function only depending on $\bar{s}\ll s$ input dimensions, then it is possible to replace $s$ with $\bar{s}$ and the transition to a faster rate will occur earlier. This is the principle of low-effective dimensionality, first introduced by Caflish et al.~in \cite{CaflishMorokoffOwen}, and is fundamental when using QMC in high dimensions. Thus, for good QMC convergence it is important to either reduce the integrand dimensionality or to order the integration variables in order of decaying importance such that the improved convergence rates can be attained.
\end{remark}

\subsection{Multilevel Monte Carlo}

The multilevel Monte Carlo method was first introduced by Heinrich in \cite{Heinrich2001} for parametric integration and subsequently described by Giles for stochastic differential equations in \cite{giles2008}. MLMC works under the assumption that one can compute a hierarchy $\{Q_\ell(\omega)\}_{\ell=1}^L$ of different approximations of the output functional $Q$ of increasing accuracy. For instance, one could consider solving \eqref{eq:generic} on a hierarchy of computational meshes of size decreasing with $\ell$. At the heart of MLMC is the expansion of $\E[Q]$ into the telescoping sum
\begin{align}
\E[Q]\approx\E[Q_L]=\sum\limits_{\ell = 1}^L\E[Q_\ell - Q_{\ell-1}],\quad Q_{0}\equiv 0.
\end{align}
Approximating each expectation in the sum on the right-hand side with standard MC then yields the MLMC estimator:
\begin{align}
\label{eq:MLMC_estimator}
\E[Q_L] \approx \hat{Q} = \sum\limits_{\ell = 1}^L\left[\frac{1}{N_\ell}\sum\limits_{n=1}^{N_\ell}(Q_\ell - Q_{\ell  - 1})(\omega^n_\ell)\right],
\end{align}
in which a key element is that the term $(Q_\ell - Q_{\ell  - 1})(\omega^n_\ell)=Q_\ell(\omega^n_\ell) - Q_{\ell  - 1}(\omega^n_\ell)$ is sampled by the same event $\omega^n_\ell$. This aspect is referred to as the \emph{MLMC coupling} and is essential for the improved performance of MLMC over MC. In fact, while the variance of each $Q_\ell$ might be large, the variance of the difference $Q_\ell-Q_{\ell-1}$ is typically much smaller due to the strong correlation between the two terms. For this reason, while the estimation of each $\E[Q_\ell-Q_{\ell-1}]$ term still occurs at a $O(N^{-1/2})$ rate, the number of samples needed to achieve a given statistical error tolerance is smaller and decreases with $\ell$. More formally, the convergence of MLMC is ensured by the following Theorem~\cite{giles2008}:

\begin{theorem}[\cite{giles2015multilevel}, Theorem $1$]
	\label{th:MLMC}
	Let $Q(\omega)$ be a random variable with finite second moment and let $Q_\ell$ be its level $\ell$ approximation. Let $Y_\ell$ be the (unbiased) MC estimator of $\E[Q_\ell - Q_{\ell-1}]$ on level $\ell$, and let $C_\ell$ and $V_\ell$ be the expected cost and variance, respectively, of each of the $N_\ell$ Monte Carlo samples needed to compute $Y_\ell$. If the estimators $Y_\ell$ are independent and there exist positive constants $\alpha$, $\beta$, $\gamma$, $c_1$, $c_2$, $c_3$, such that $\alpha\geq \frac{1}{2}\min(\beta,\gamma)$ and
	\begin{align}
	\label{eq:MLMC_theorem_bounds}
	|\E[Q_\ell - Q]|\leq c_12^{-\alpha\ell},\qquad V_\ell\leq c_2 2^{-\beta\ell},\qquad C_\ell\leq c_32^{\gamma\ell},
	\end{align}
	then there exist a positive constant $c_4$ such that, for all $\varepsilon < e^{-1}$, there is a level number $L$ and number of samples $N_\ell$, such that the mean square error (MSE) of the MLMC estimator
	$\hat{Q}=\sum\limits_{\ell=1}^L Q_\ell$ is bounded:
	\begin{align}
	  \rm{MSE} = \E[(\hat{Q} - \E[Q])^2]\leq \varepsilon^2,
	\end{align}
	and its total computational complexity is bounded:
	\begin{align}
	\E[C_{tot}] \leq \left\{\begin{array}{lr} c_4\varepsilon^{-2}, & \beta > \gamma,\\c_4\varepsilon^{-2}(\log\varepsilon)^2, & \beta = \gamma,\\c_4\varepsilon^{-2 - (\gamma - \beta)/\alpha}, & \beta < \gamma.\end{array}\right.
	\end{align}
\end{theorem}

\begin{remark}
  The MLMC parameters $\alpha$, $\beta$, and $\gamma$ must be estimated if not known \emph{a priori}. However, for PDE applications, as it is the case here, \emph{a priori} error estimates are typically available. The optimal number of samples on each level $N_\ell$ and the maximum level $L$ needed can similarly be estimated.
\end{remark}
\begin{remark}
  \label{rem:max_level_bounded}
  Theorem \ref{th:MLMC} assumes that one can increase the maximum level $L$ at will. In practical applications this might not always be possible, e.g.~when the computational resources available are limited.
\end{remark}

In this study, we modify the original MLMC algorithm by weighting the relative importance of bias and statistical error as introduced by Haji-Ali et al~\cite{haji2016optimization}. The MSE of the MLMC estimator is given by
\begin{align}
MSE = \hat{V} + \E[\hat{Q} - Q]^2,
\end{align}
where $\hat{Q}$ is the MLMC estimator with variance $\hat{V}$. To ensure that $MSE\leq\varepsilon^2$, we enforce the bounds,
\begin{align}
\label{eq:theta}
\hat{V}\leq (1-\theta) \varepsilon^2,\quad \E[\hat{Q} - Q]^2 \leq \theta \varepsilon^2,
\end{align}
where $\theta\in (0,1)$ is a weight balancing the two terms so as to obtain comparable error reduction. Small values of $\theta$ reduce the number of samples needed and are therefore preferred when the bias is small. Conversely, large values are beneficial when the bias is large as they allow to achieve smaller tolerances without adding finer levels to the hierarchy.
 
\subsection{Gaussian field sampling techniques}

When $Q(\omega)$ depends on a random field (e.g.~through the coefficients of \eqref{eq:generic}), an efficient sampling technique is also needed as part of any Monte Carlo method. Unfortunately, sampling generic random fields from a given distribution is normally a prohibitive task. In the specific case in which the field to be sampled is Gaussian, however, the sampling problem becomes tractable, albeit still computationally expensive. Different Gaussian field sampling methods are available in the literature. The most common are: 1) Factorization of the covariance matrix, including Hierarchical matrix approximation \cite{FeischlKuoSloan2018,Khoromskij2009,Hackbush2015HMatrices,DolzHarbrechtSchwab2017} 2) Karhunen-Lo\`eve expansion of the random field (cf.~Section 11.1 in \cite{sullivan2015introduction} and e.g.~\cite{rodriguez2019uncertainty}) and 3) the circulant embedding method which uses FFT \cite{WoodChan1994,dietrich1997fast,GrahamKuoEtAl2018,BachmayrGraham2019}. 

In this paper we use an alternative method known as the stochastic PDE (SPDE) approach for the sake of efficiency \cite{Whittle1954,Lindgren2011,Croci2018,CrociMLQMC}. This sampling strategy is based on approximately drawing realisations of a Mat\'ern field by solving the following elliptic PDE \cite{Whittle1954,Lindgren2011}:
\begin{align}
\label{eq:white_noise_eqn}
\left(I - \kappa^{-2}\Delta\right)^{k}u(\bm{x},\omega) = \eta \W(\cdot, \omega),\quad \bm{x}\in \hat{G}\subseteq\R^d,\quad \omega\in\Omega,\quad \nu = 2k - d/2 > 0,
\end{align}
where $\nu$ is the smoothness parameter of the Mat\'ern covariance to be sampled (cf.~equation \eqref{eq:Matern}), $\W$ is spatial Gaussian white noise in $\R^d$, $d\leq 3$, $k>d/4$, and the equality has to hold almost surely and be interpreted in the sense of distributions. When $k$ is an integer, as we shall assume here, solving \eqref{eq:white_noise_eqn} is equivalent to solving a second-order PDE $k$ times.  The constant $\eta>0$ is a scaling factor that depends on the Mat\'ern covariance parameters $\sigma$, $\lambda$ and $\nu$, cf.~\cite{Croci2018}. The solution to $u(\bm{x},\omega)$ is a Mat\'ern field only if $\hat{G}\equiv \R^d$. Otherwise, for general $\hat{G}\subset\R^d$, $u$ is still a Gaussian field, but its covariance is only approximately Mat\'ern, with the error in the covariance decaying exponentially away from the boundary of $\hat{G}$ \cite{Khristenko2018}. For this reason in this paper we will always assume that the field sample is needed on a domain $G\subset \hat{G}$ and that we solve \eqref{eq:white_noise_eqn} on a domain $\hat{G}$ large enough so that this source of error is negligible over $G$. We impose homogeneous Dirichlet boundary conditions on $\partial \hat{G}$.

Note that the term $\W$ is random and for each sample of $u$ needed, we must sample $\W$ from its distribution and solve~\eqref{eq:white_noise_eqn}. The sampling of $\W$ is non-trivial in itself, especially when samples are needed within a MLMC or QMC framework: in the QMC case, a good variable ordering is required to achieve good convergence with respect to the number of samples while in the MLMC case the multilevel coupling must be enforced correctly to obtain a considerable variance reduction. In this paper, we use the sampling techniques developed in \cite{Croci2018} for the MLMC case and in \cite{CrociMLQMC} for the (ML)QMC case to address these requirements. Both these MLMC and (ML)QMC sampling strategies are efficient with optimal cost complexity, making the sampling of $u$ a linear cost and memory operation. This optimal complexity is especially advantageous when dealing with physiologically relevant applications with complex geometries in 3D.

\section{A stochastic model of tracer transport in brain tissue}
\label{sec:methods}

This paper aims to evaluate the numerical and computational performance of different Monte Carlo methods for quantifying uncertainty in stochastic models of tracer transport in brain tissue. In particular, we focus on two comprehensive coefficient models with random diffusion and velocity fields. In addition to reflecting existing hypotheses on ISF tracer transport, these models are designed to offer a suitable challenge to the MLMC and QMC algorithms presented in \cite{Croci2018} and \cite{CrociMLQMC}.

\subsection{The ISF tracer transport equation}

\begin{figure}[!h]
	\begin{center}
		\includegraphics[width=\linewidth]{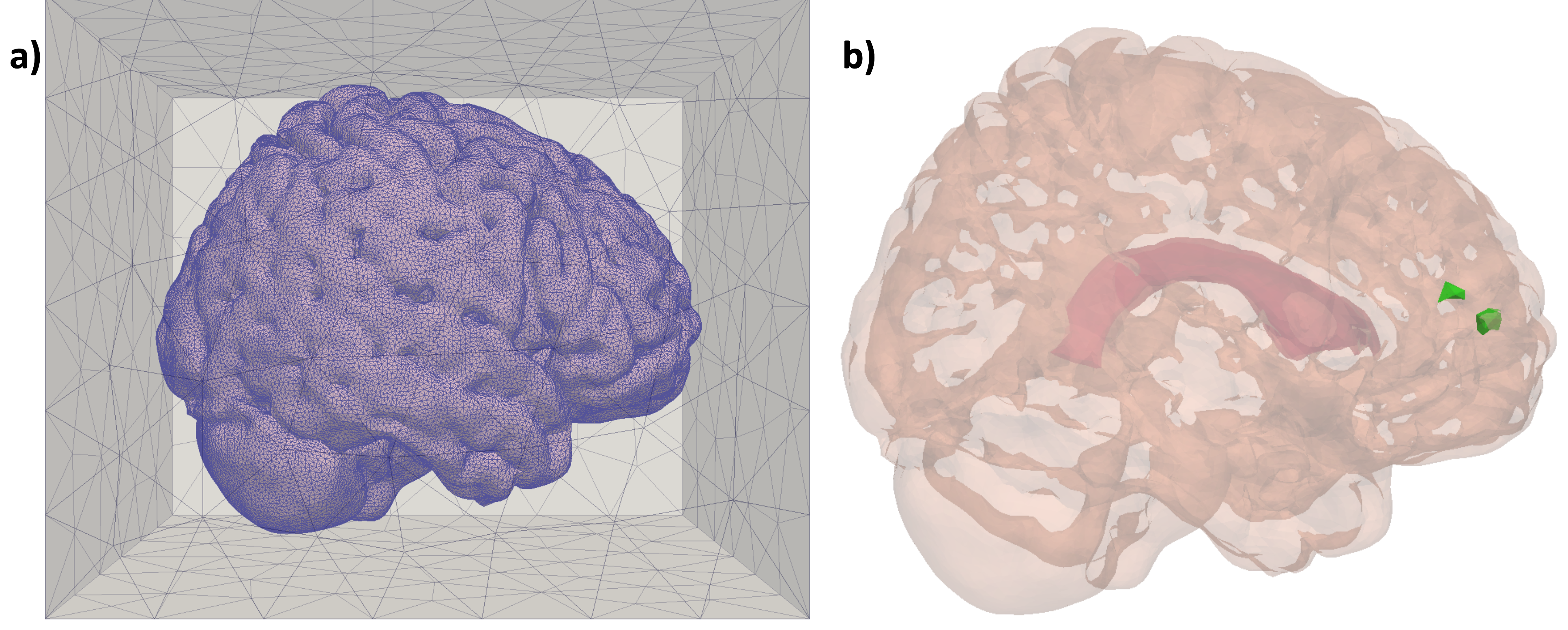}
		\caption{\textit{The computational domains and points of interest. a) The coarsest computational mesh, embedded in the auxiliary box used to sample the Mat\'ern fields. b) The domain representing the brain parenchyma. The lateral ventricles are shown in red, while the parenchyma is pink. Two smaller regions of interest are marked in green: The leftmost green region, $S_w$, is within the white matter, while the rightmost green region, $S_g$, is within the gray matter.}}
		\label{fig:1}
	\end{center}
\end{figure}

We begin by considering the baseline model derived by the authors in \cite{CrociVinjeRognes2019}, and briefly introduced in Section~\ref{sec:preliminaries}, describing the evolution of tracer concentration within the brain parenchyma under uncertainty: find the tracer concentration $c = c(t, \bm{x}, \omega)$
for $\bm{x} \in G$, $\omega \in \Omega$ and $t \geq 0$ such that
\begin{equation}
\label{eq:transport}
\dot c(t,\bm{x}, \omega)
+ \nabla \cdot (\bm{v}(\bm{x},\omega) c(t,\bm{x},\omega))
- \nabla \cdot (D^{\ast}(\bm{x},\omega) \nabla c(t,\bm{x},\omega)) + rc(t,\bm{x},\omega)= 0.
\end{equation}
As before, $G\subset\R^3$ is the brain parenchyma domain comprised of white and gray matter from the Colin27 human adult brain atlas FEM mesh~\cite{FangEtAl2010} version 2 (Figure~\ref{fig:1}).
The superimposed dot represents the time derivative, $D^{\ast}$
is the effective diffusion coefficient of the tracer in the tissue
(depending on the tracer free diffusion coefficient and the tissue
tortuosity) \cite{Nicholson2001}, $\bm{v}$ represents a convective fluid
velocity and $r \geq 0$ is a drainage coefficient potentially
representing e.g.~capillary absorption~\cite{orevskovic2017new} or
direct outflow to lymph nodes~\cite{RingstadEtAl2017}. The parenchymal domain
is assumed to not contain any tracer initially: $c(0, \bm{x}, \omega) = 0$. This model reflects the MRI-study of Ringstad et al.~\cite{RingstadEtAl2017} in which 0.5 mL of 1.0 mmol/mL of the radioactive tracer gadobutrol was injected in the spinal canal (intrathecally) of 15 hydrocephalus patients and eight reference subjects. 

\subsection{Boundary conditions}

Let the brain boundary $\partial G = \partial G_S \cup \partial G_V$,
with $\partial G_S$ representing the interface between the
brain parenchyma and the subarachnoid space (SAS), and $\partial
G_V$ representing the interface between the brain parenchyma
and the cerebral ventricles, respectively. We assume the tracer concentration on the SAS interface to be known and impose no ventricular outflux. As boundary conditions for~\eqref{eq:transport}, we thus prescribe
\begin{align}
\label{eq:DirichletBC}
c = g(t,\bm{x}) \text{ on } \partial G_S, \\
\label{eq:NeumannBC}
D^{\ast}\nabla c \cdot \bm{n} = 0 \text{ on } \partial G_V,
\end{align}
where $\bm{n}$ is the unit normal vector pointing outward from $\partial G$. The function $g(t,\bm{x})$ models the movement of
tracer starting from the lower cranial SAS and traveling upward in the
CSF surrounding the brain as observed in the study by Ringstad et
al.~\cite{RingstadEtAl2017}, and the form
\begin{equation}
\begin{split}
\label{eq:Dirichlet}
g(t, \bm{x}) &= c_{\rm CSF}(t) \, h(t, \bm{x}), \\
h(t, \bm{x}) &= \left(0.5+\frac{1}{\pi}\arctan(-a(x_3-z_0-\upsilon_zt))\right),
\end{split}
\end{equation}
where $\bm{x} = [x_1, x_2, x_3]$ and $c_{\rm CSF}(t)$ is the
average tracer concentration in the SAS,
while $h(t,\bm{x})$ represents its spatial distribution. The variable $\upsilon_z = 1.5 \times 10^{-5}$ m/sec is the speed of
tracer movement upwards in the SAS, while $a = 20$ m$^{-1}$ reflects the gradient of
tracer concentration from the lower to the upper cranial SAS. The value $z_0 =
-0.2$ m is the initial distance from the lateral ventricles reached by the tracer at $t=0$.

The average SAS tracer concentration $c_{\rm CSF}$ is modelled as follows. Let $n_0 = 0.5$ mmol be the total amount of tracer initially
injected in the CSF and let $V_\text{CSF} = 140$ mL be the total CSF volume in the human SAS and ventricles \cite{Wood2013}. Then, the average concentration in the SAS right after injection is given by $c_{\rm CSF}(0)$ = 0.5 mmol/140 mL = 3.57 mol/m$^3$. Assuming conservation of tracer molecules, the total amount of tracer in the brain and in the SAS
plus or minus the tracer otherwise absorbed or produced is constant in time,
and is equal to the initial amount $n_0$. This observation gives the approximate relation
\begin{equation}
\int_G \bar{c}(t,\bm{x}) \dx
+ c_{\rm CSF}(t) V_{\rm CSF}
+ \int_0^t \int_G r \bar{c}(\tau,\bm{x}) \dx \, \mathrm{d}\tau = n_0.
\end{equation}
where, for simplicity, $\bar{c}$ is given by a deterministic solution of \eqref{eq:transport} with boundary conditions \eqref{eq:Dirichlet} in which all the random coefficients are replaced by their average. Solving for $c_{\rm CSF}$, we finally let
\begin{align}
\label{eq:csf-conc}
c_{\rm CSF}(t)
= \frac{1}{V_{\rm CSF}} \left (n_0 - \int_{G} \bar{c}(t,\bm{x}) \text{ d}\bm{x}
- \int_0^t \int_G r \bar{c}(\tau,\bm{x}) \text{ d}\bm{x} \, \mathrm{d} \tau \right).
\end{align}

\subsection{Quantities of interest}

We consider different output quantities (functionals) describing the characteristics
of tracer movement within the brain. For each time $\tau=30k$ min for $k=1,\dots, 48$ (from half an hour from injection to one day after), we consider the
total amount of tracer in the gray matter $Q_g$ and in the
white matter $Q_w$:
\begin{align}
Q_g(\tau, \omega) = \int_{D_g} c(\tau, \bm{x}, \omega) \text{ d}\bm{x}, \quad
Q_w(\tau, \omega) = \int_{D_w} c(\tau, \bm{x}, \omega) \text{ d}\bm{x}.
\label{eq:Q_gw}
\end{align}
Additionally, we consider two localized
concentration measures: the average tracer concentrations
$q_g$ and $q_w$ in two smaller regions, one within the gray matter $S_g$ and one within the white matter $S_w$ respectively:
\begin{equation}
q_{g}(\tau, \omega) = \frac{1}{V_g} \int_{S_g} c(\tau, \bm{x},\omega) \text{ d}\bm{x}, \quad
q_{w}(\tau, \omega) = \frac{1}{V_w} \int_{S_w} c(\tau, \bm{x},\omega) \text{ d}\bm{x},
\label{eq:q_gw}
\end{equation}
where $V_g$ and $V_w$ is the volume of the gray and white matter
subregions, respectively. The size and location of these
subregions are shown in Figure~\ref{fig:1}b.

\section{Coefficient models}
\label{sec:coefficient_models}

The effective diffusion coefficient of a solute (or tracer) and the velocity field are heterogeneous \cite{tuch2002high,JessenEtAl2015,KiviniemiEtAl2016,rajna20193d} within the parenchyma and also vary from individual to individual. To account for both these types of variation and for the uncertainty in the coefficient magnitude we model them as random variables or fields.

\subsection{Diffusion coefficient}
\begin{figure}[!h]
	\begin{center}
		\includegraphics[width=\linewidth]{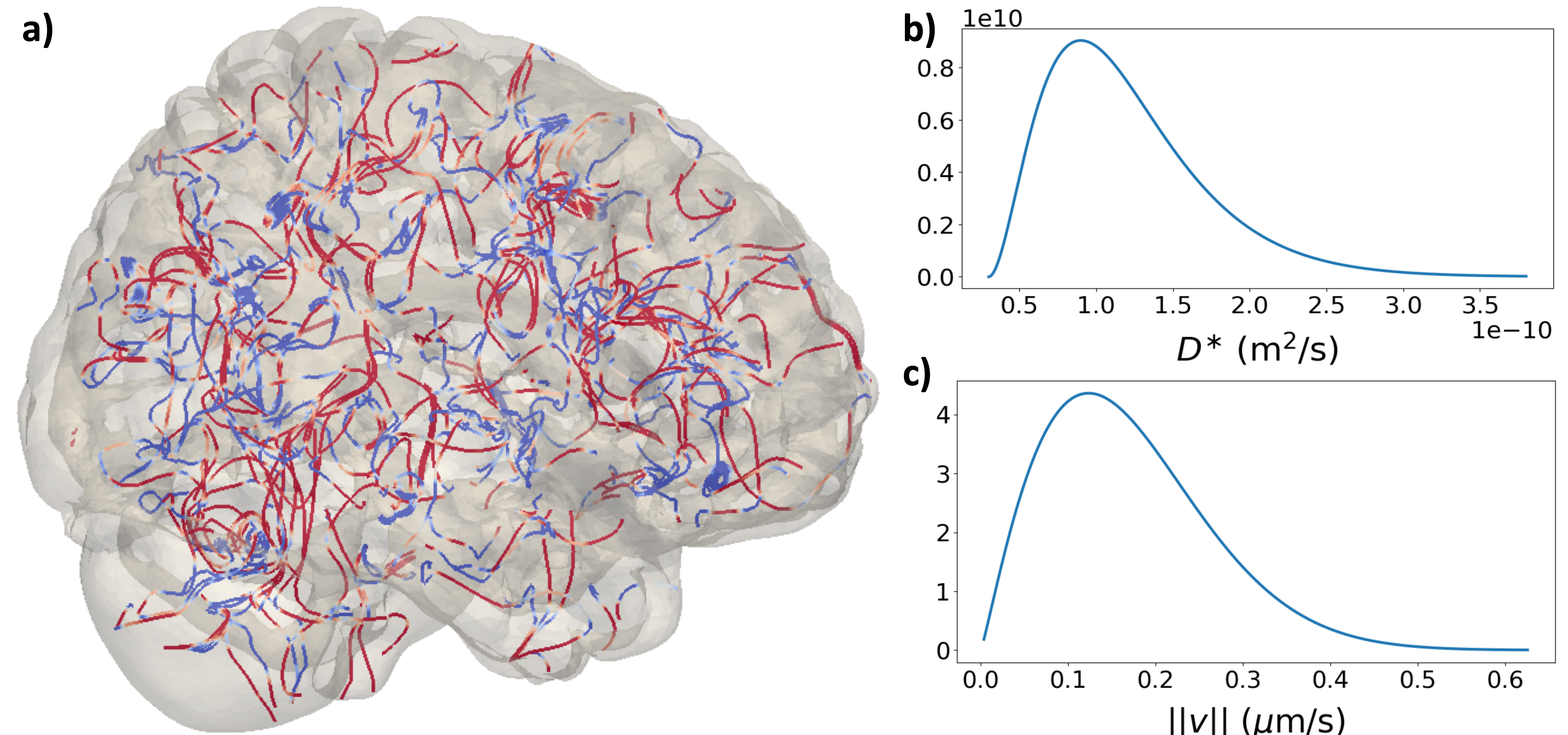}
		\caption{\textit{Stochastic aspects of diffusion and velocity fields. a) Streamlines of the velocity field $\bm{v}_{\rm base}$, representing a random distribution of blood vessels. Colors indicate velocity magnitude, and an arbitrary small scaling range is chosen for visual purposes. b) Probability density of the diffusion coefficient $D^\ast$. c) Probability density of the glymphatic circulation velocity magnitude $| \bm{v} |_2$ cf.~\eqref{eq:velocity_model1}.}}
		\label{fig:2}
	\end{center}
\end{figure}

Let $D^{\ast}_{\rm Gad} = 1.2 \times 10^{-10}$ m/s$^2$ be the average
parenchymal gadobutrol diffusivity \cite{RingstadEtAl2018}. We model the effective diffusion coefficient as
\begin{equation}
  \label{eq:diffusion}
  D^{\ast}(\bm{x},\omega) = 0.25 \times D^{\ast}_{\rm Gad} + D^{\ast}_{\gamma} (\bm{x}, \omega),
\end{equation}
where $D^{\ast}_{\gamma}$ is a random field such that for each fixed $\bm{x}
\in G$, $D^{\ast}_{\gamma}(\bm{x}, \omega)$ is a gamma-distributed
random variable with shape $k=3$
and scale $\theta = 0.75 \times D^{\ast}_{\rm Gad}/k$. This choice of parameters ensures the positivity of $D^{\ast}$ with probability $1$. Furthermore, it reflects the average value and variability reported in the literature, namely we have $\E[D^{\ast}]=D^{\ast}_{\rm Gad}$, with values larger than $3$ times the average being attained with very low probability \cite{Nicholson2001,RingstadEtAl2018}. The probability distribution of $D^{\ast}(\bm{x},\cdot)$ is shown in Figure~\ref{fig:2}b.

To sample $D^{\ast}_{\gamma}$ from its distribution, we
first sample a Mat\'ern field $X(\bm{x},\omega)$
using the techniques presented in \cite{Croci2018} and \cite{CrociMLQMC},
and then transform it into a gamma random field by using a
copula~\cite{nelsen2007introduction}. This consists in setting
$D^{\ast}_{\gamma}(\bm{x},\omega)=F^{-1}(\Phi(X(\bm{x},\omega)))$, where
$F^{-1}$ is the inverse cumulative density function (CDF) of the
target (gamma) distribution, $\Phi$ is the CDF of the standard normal
distribution and $X(\bm{x},\omega)$ is a standard (zero mean, unit
variance) Mat\'ern field with smoothness parameter $\nu=2.5$ and
correlation length $\lambda = 0.01$ m, cf.~\eqref{eq:Matern}. 
Note that $\Phi$ maps any standard normal random variable to a
standard uniform random variable and that $F^{-1}$ maps any standard
uniform random variable to the target distribution, hence the function
$F^{-1}(\Phi(x))$ maps standard random variables to the target gamma
distribution. Samples of $D^{\ast}_{\gamma}(\bm{x},\omega)$ obtained this
way will preserve the same Spearman correlation and smoothness
properties of $X(\bm{x},\omega)$, but will present a different covariance
structure \cite{nelsen2007introduction}.

\subsection{Velocity and drainage coefficients}

We now turn to define two models (Model 1, 2) for brain tissue fluid
movement and clearance. Both models are combined with the random
diffusion field defined by~\eqref{eq:diffusion}. For further
physiological considerations, see e.g.~\cite{CrociVinjeRognes2019}.

\subsubsection{Model 1 - Glymphatic velocity model with directionality}

For Model 1, we model fluid transport along paravascular spaces in direct communication with the SAS~\cite{JessenEtAl2015} under the following assumptions. 1) Substantial changes in the velocity field occurs at a distance proportional to a characteristic distance (correlation length $\lambda$) between an arteriole and a venule. 2) The velocity field can be represented as a sum of a glymphatic velocity field associated with arterioles and venules and a velocity field represents a directional movement due to larger blood vessel structures. 3) Almost no fluid is filtrated or absorbed by capillaries, thus $\bm{v}_{\rm base}$ is divergence-free, while $\bm{v}_{\rm dir} $ only has a small net flow of 0.007 mL/min out of the parenchyma. We then let the drainage term $r=0$ in \eqref{eq:transport} and define the stochastic velocity field
\begin{equation}
\bm{v}(\bm{x}, \omega) = \bm{v}_{\rm base}(\bm{x}, \omega) + \bm{v}_{\rm dir}(\bm{x}).
\end{equation}

The stochastic glymphatic velocity field $\bm{v}_{\rm base}$ is given by
\begin{align} 
\label{eq:velocity_model1}
\bm{v}_{\rm base}(\bm{x},\omega) = v_{\text{avg}}\cdot \bar{\eta}(\lambda)\ \sqrt{U(\omega)} \left(\nabla \times \left[\begin{array}{c} X(\bm{x},\omega) \\
Y(\bm{x},\omega) \\
Z(\bm{x},\omega)
\end{array}\right]\right),
\end{align}
where $\bar{\eta}(\lambda) = \lambda/\sqrt{8}$ is a scaling
constant chosen such that the magnitude of $\bm{v}$ (denoted $|\bm{v}|$) satisfies
$\E[|\bm{v}|^2]^{1/2} = v_{\text{avg}}$, $U(\omega)$ is an independent standard uniform
random variable and $X(\bm{x},\omega)$, $Y(\bm{x},\omega)$ and $Z(\bm{x},\omega)$
are standard i.i.d.~Mat\'ern fields with
$\nu=2.5$ and correlation length $\lambda = 1020\, \mu$m. Taking the curl of the random vector field $[X,Y,Z]^T$ ensures that $\bm{v}_{\rm base} $ is divergence free. A sample of the streamlines of $\bm{v}_{\rm base}$ is shown in Figure~\ref{fig:2}a while its velocity magnitude distribution is shown in Figure~\ref{fig:2}c. The deterministic second term $\bm{v}_{\rm dir}$ represents a directional velocity field induced by large vascular structures \cite{CrociVinjeRognes2019} and is given by 
\begin{equation}
\bm{v}_{\rm dir}(\bm{x}) = -v_f 
\left [
\begin{array}{c}
\arctan(15x_1)(|x_1|-0.1) \\
\arctan(15x_2)(|x_2|-0.1) \\
-0.9x_3+0.06-\sqrt{x_1^2+x_2^2}
\end{array}
\right ],
\label{eq:model-V2}
\end{equation}
where $v_f = 2 \times 10^{-6}$ m/s. 

\begin{remark}
We now briefly show that $\E[\bm{v}_{\rm base}] = \bm{0}$ and hence $\E[\bm{v}]=\bm{v}_{\rm dir}(\bm{x})$. In fact, note that the partial derivative $\partial X/\partial x_i$ of a zero-mean Gaussian field $X(\bm{x},\omega)$
with a twice differentiable covariance $C(\bm{x},\bm{y})$ is still
a zero-mean Gaussian field with covariance given by $\partial^2 C(\bm{x},\bm{y})/(\partial x_i \partial y_i)$ (cf.~Section 2.3 in \cite{PetterAbrahamsen1997}).
The curl components of the field within the brackets in \eqref{eq:velocity_model1} are therefore just sums of independent Gaussian fields, and hence Gaussian as well. Similarly, this also shows that the covariance of $\bm{v}$ has the same correlation length as the Mat\'ern fields $X$, $Y$ and $Z$ since the second derivative of a Mat\'ern covariance (cf.~\eqref{eq:Matern}) has the same correlation length as the original covariance function, although it is not Mat\'ern anymore.
\end{remark}

\subsubsection{Model 2 - Capillary filtration model with arterial inflow and sink term}

For Model 2, we consider an alternative velocity representation in which CSF enters the brain parenchyma along spaces surrounding penetrating arteries~\cite{mestre2018flow,JessenEtAl2015,albargothy2018convective,BedussiEtAl2017}. In this case, the velocity field is taken to be
\begin{align}
\bm{v}(\bm{x}, \omega) = \bar{v}(\omega)\exp\left(-\frac{3(R-||\bm{x}-\bm{x}_c||)^2}{R^2-(R-||\bm{x}-\bm{x}_c||)^2}\right)(\bm{x}_c-\bm{x}),
\label{eq:v3-velocity}
\end{align}
representing a net flow of CSF into the brain. The flow field is radially symmetric and directed inwards from the outer SAS to a spherical region of radius $R = 8$ cm around a center point $\bm{x}_c$ within the lateral ventricles. Here $\bar{v}(\omega)$ is a gamma-distributed random variable chosen such that the probability distribution of the velocity magnitude is comparable
to that of Model 1. The shape parameter is $k=2$ and the scale parameter is set such that again $\E[|\bm{v}|^2]^{1/2} = v_{\rm avg}$. Note that in this model $\E[\bm{v}]\neq \bm{0}$ and the main source of uncertainty is the random variable ($\bar{v}(\omega)$) rather than the spatially dependent random field.

Finally, we set a non-zero sink coefficient of $r = 1 \times 10^{-5}$ s$^{-1}$, to model the assumption that ISF is drained along some alternative outflux route within the brain parenchyma. The value of $r$ chosen corresponds to a $40\%$ drainage of the injected tracer over $48$ hours.

\section{Numerical solution of the stochastic models}
\label{sec:numerical_details_stochastic_models}

More advanced Monte Carlo methods, such as QMC, MLMC and MLQMC, are known to outperform standard Monte Carlo methods, given an appropriate problem setting. In particular, QMC requires either the output functional to be of low effective dimensionality with respect to the random input, or the input dimensions to be ordered in order of decaying importance \cite{CaflishMorokoffOwen}. Even though our QMC method~\cite{CrociMLQMC} is designed to expose the leading-order dimensions in each random field input, hence providing a suitable variable ordering, it is not known whether this strategy would prove to be effective for the problem at hand. In fact, two main challenges arise: 1) The state equation \eqref{eq:transport} must be solved on a complicated 3D geometry, and 2) the random input dimensionality is large. The latter point is especially relevant in connection with Model 1 in which $4$ infinite-dimensional random fields appear as coefficients. Both these challenges significantly increase the input dimensionality, possibly affecting QMC performance.

On the other hand, MLMC brings a different set of performance requirements. A good MLMC coupling is ensured by our coupling technique~\cite{Croci2018}, but the technique also hinges on the construction of an appropriate mesh hierarchy on which a good rate of decay of bias and variance can be appreciated. Even though such a hierarchy can always be obtained by increasing the mesh refinement in theory, the availability of computational resources and time may limit the maximal refinement level in practice. As it is \emph{a priori} unclear whether QMC and/or MLMC actually bring any significant advantages with respect to standard MC when solving \eqref{eq:transport}, we here evaluate both algorithms to determine which if any method performs the better.

We refer the reader to the book by Lemieux \cite{Lemieux2009} and to the review article by Giles \cite{giles2015multilevel} for further information about QMC and MLMC methods. In what follows, we detail the numerical approach adopted for the solution of \eqref{eq:transport}.

\subsection{Weak form and discretisation}

We solve \eqref{eq:transport} numerically using the finite element
method (FEM).
Let $H^1_{S}(G)$ be the standard Sobolev space of weakly
differentiable functions that are zero on the boundary $S$~\cite{Evans2010}. For $c, s \in H^1(G)$, we define
\begin{align}
a(c,s) = \inner{\Div (\bm{v}c)}{s} + \inner{D^\ast \nabla c}{\nabla s} + \inner{rc}{s},
\end{align}
where by $\inner{\cdot}{\cdot}$ we indicate the standard $L^2(G)$ inner product:
\begin{equation}
  \inner{c}{s} = \int_{G} c s \dx.
\end{equation}
After a (second-order) implicit mid-point time discretisation with time step $\Delta t$, the continuous weak form of \eqref{eq:transport} reads: find $c^n\in H^1(G)$ such that for all $s\in H^1_S(G)$ and a.s.,
\begin{align}
  \label{eq:time_discretisation_transport}
  \inner{c^n - c^{n-1}}{s} + \frac{\Delta t}{2}(a(c^n, s) + a(c^{n-1}, s))= 0, \\
  c^0 \equiv 0,
  \quad c^n = c_{\text{CSF}}^n h(t^n,\bm{x}) \text{ on } \partial G_S,
\end{align}
where $c^n$ thus represents an approximation to $c(t^n,\cdot)$ with $t^n = n\Delta t$ for $n=0,\dots,n_T-1$ and $n_T-1=T/(\Delta t)$ and $c_{\text{CSF}}^n$ is an approximation of $c_{\text{CSF}}(t^n)$, defined in \eqref{eq:csf-conc}. We approximate $c_{\text{CSF}}^n$ by approximating the time integral in \eqref{eq:csf-conc} with the trapezoidal rule:
\begin{equation}
\begin{split}
\label{eq:csf-conc-discretized}
c_{\rm CSF}(t^n)\approx c_{\text{CSF}}^n
= \frac{1}{V_{\rm CSF}}  \left (n_0 - \int_{G} \bar{c}^n \text{ d}\bm{x} \right. \left.- \frac{\Delta t}{2} \left(2\sum\limits_{i=1}^{n-1}\int_G  r \bar{c}^i \text{ d}\bm{x}  + \int_G  r \bar{c}^n \text{ d}\bm{x}\right) \right).
\end{split}
\end{equation}
We note that the expression on the right-hand side is known as $\bar{c}$ can be pre-computed once. This decoupling results in a second-order scheme in time for $c$. To compute $\bar{c}$ we adopt the same discretisation, but we approximate $c_{\text{CSF}}^n$ explicitly (by replacing $n$ with $n-1$ in the right-hand side of \eqref{eq:csf-conc-discretized}) thus avoiding the non-local, implicit boundary condition. This approximation gives to a first-order scheme for $\bar{c}$, for which we compensate by computing the approximation using a very small time step ($\Delta t_{\bar{c}} = 30\times 2^{-6}$ min) and we solve for $\bar{c}$ on the finest mesh available (see later in this section for a description of the meshes and time step sizes used).

We discretize \eqref{eq:time_discretisation_transport} in space using the FEM. Given a FEM approximation subspace $V_h\subset H^1_S(G)$ , the fully discrete weak form of \eqref{eq:transport} reads: find $c^n_h \in V_h$ such that, for all $s \in V_h$ and a.s.,
\begin{align}
\label{eq:transport_fully_discrete}
  \inner{c_h^n - c_h^{n-1}}{s} + \frac{\Delta t}{2}(a(c_h^n, s) + a(c_h^{n-1}, s))= 0, \\
  c_h^0 \equiv 0,
  \quad c^n_h=c_{\text{CSF}}^{n,h}h(t^n,\bm{x})\text{ on } \partial G_S,
\end{align}
where $c_{\text{CSF}}^{n,h}$ is given by \eqref{eq:csf-conc-discretized} in which $c^{n-1}$ and $c^i$ are replaced by $c^{n-1}_h$ and $c^i_h$ respectively. We let $V_h$ be composed of piecewise linear continuous Lagrange elements defined relative to a mesh $\mathcal{T}_h$ of $G$ of mesh resolution $h$.

\subsection{Meshes and time steps}

We discretize the domain $G$ by using various refinements of the Colin27 human adult brain atlas simplicial mesh~\cite{FangEtAl2010} (version 2). We construct a multilevel hierarchy in which the coarsest level is given by one uniform refinement of the original brain mesh and the other $2$ levels are obtained through uniform refinement. On level $\ell$, we fix $(\Delta t)_\ell = 15\times 2^{-\ell}$ min and we terminate the simulations after $T=1$ day. The numbers of cells and vertices of the meshes in the hierarchy are given in Table \ref{tab:hierarchy_meshes}. The Mat\'ern fields are sampled on an extended domain, i.e.~a mesh of a larger box domain $\hat{G}$ of size sufficiently large to make the domain truncation error negligible (dimensions $0.16 \times 0.21 \times 0.17$ m$^3$) \cite{Khristenko2018}, and then restricted to the brain mesh. The outer box together with the embedded Colin27 mesh is shown in Figure~\ref{fig:1}a. Each outer box mesh is constructed with the meshing software Gmsh \cite{gmsh} (dev version 4.2.3-git-023542a) such that the corresponding brain mesh is nested within. Furthermore, the box meshes are graded such that the cell size gradually gets larger away from the brain boundary (Mat\'ern field values are only needed in the brain domain).

\begin{table}[h!]
	\centering
	\begin{tabular}{@{}c|cccc@{}}
		\toprule
		& Colin27    & $\ell=1$      & $\ell=2$       & $\ell=3$        \\ \midrule
		n cells    & $249\,361$ & $1\,994\,888$ & $15\,959\,104$ & $127\,672\,832$ \\
		n vertices & $55\,066$  & $391\,559$    & $2\,905\,985$  & $22\,282\,705$  \\ \bottomrule
	\end{tabular}
	\caption{Number of cells and vertices of the Colin27 mesh and of the meshes of the MLMC hierarchy.}
	\label{tab:hierarchy_meshes}
\end{table}

\subsection{Numerical stability considerations}
\label{sec:Peclet}

For the parameter regimes considered here, the problem~\eqref{eq:transport} is only mildly convection-dominated, with an upper estimate of the P\'eclet number of 
\begin{align}
Pe \approx \frac{\hat{L}  v_{\rm avg}}{D^{\ast}_{\rm Gad}} = O(10^2),
\end{align}
where $\hat{L} \approx 0.084$ m is half the diameter of the computational
domain, $v_{\rm avg} = 0.17 \mu$m/s, and $D^{\ast}_{\rm Gad} = 1.2
\times 10^{-10}$ m/s$^2$. Given the fine computational meshes, we obtain low-probability worst-case cell P\'eclet numbers of $\approx 43\times 2^{-\ell}$ on level $\ell$ of the MLMC hierarchy used. In the numerical experiments, convection-related numerical instabilities were not observed.

However, in initial investigations, we observed that the FEM solution undershoots near the boundary, attaining negative concentration values. This is a known phenomenon in the literature and it does not depend on the velocity field, but it is typical of diffusion problems with Dirichlet-type boundary conditions \cite{thomee2014positivity}. We address this problem by a mass-lumping technique, which is known to reduce this effect \cite{thomee2014positivity}. This ill-behaviour disappears as the mesh is refined to the extent that no undershoots were observed on the finer levels of the MLMC hierarchy. In our numerical experiments, MLMC convergence was resilient to non-physical behaviour on the coarser levels. In fact, we noted that these coarse level solutions would still act as a good control variate for the finer levels. 

\subsection{Solver and software}
\label{sec:solver_software_limitedCPU}

For the computations, we combined the University of Oxford Mathematical Institute computing servers and the University of Oslo Abel supercomputing cluster. The total amount of CPU hours (CPUh) available on the Abel cluster for the project were approximately $400\,000$ CPUh with the rule that computations with high memory requirements ($>4$ GB) would cost an amount of CPU hours given by the formula
\begin{align}
\text{CPU hours cost}\quad = \quad \text{n hours}\quad \times\quad (\text{memory per CPU})/(4 \text{ GB}).
\end{align}
The computation was memory bound, therefore causing the overall computational resources to be limited. We used the Abel cluster exclusively for the samples on the finest MLMC level.

We used the FEniCS FEM software \cite{LoggEtAl2012}. The linear systems of equations were solved using the PETSc~\cite{balay2014petsc} implementation of the GMRES algorithm preconditioned with the BoomerAMG algebraic multigrid algorithm from Hypre \cite{hypre}. We use the MLMC and QMC routines from the open-source software FEMLMC \cite{femlmc}, which contains the implementation of the algorithms presented in \cite{Croci2018} and \cite{CrociMLQMC}. For the FEMLMC Mat\'ern field sampling solver, we declare convergence when the absolute size of the preconditioned residual norm is below a tolerance of $10^{-8}$. We use the same stopping criterion for the GMRES solver.

Convergence of the numerical solver was verified with a convergence test comparing different mesh refinements and time steps for a set of deterministic
worst-case models (with large velocities and small diffusion coefficients), see e.g.~\cite{CrociVinjeRognes2019} (Supplementary Material).

\subsection{QMC and MLMC algorithms}
\label{sec:QMC_MLMC_algorithms}

We now describe the QMC and MLMC algorithms used more in details. We adopt the following (standard) randomized QMC algorithm \cite{Lemieux2009}:
\paragraph{QMC algorithm}
\begin{enumerate}
	\item Set the required tolerance $\varepsilon$, $\theta\in(0,1)$. Set the mesh size $h$ and $\Delta t$ to ensure that the bias estimate is lower than $\sqrt{\theta}\varepsilon$;\vspace{-3pt}
	\item Get an initial estimate of $\hat{V}_{\text{QMC}} = \V[I_N(\omega)]$ (cf.~\eqref{eq:randomized_QMC}) with $N=1$ samples and $M=32$ randomizations;\vspace{-3pt}
	\item While $\hat{V}_{\text{QMC}} > (1-\theta)\varepsilon^2$, double $N$.\vspace{-3pt}
\end{enumerate}
The total QMC cost is given by $C_{\text{tot}}=\bar{C}NM$, where $N$ is the final number of samples taken and $\bar{C}$ is the (expected) computational cost of computing one QMC sample.
\begin{remark}
	In the random PDE case, typically the bias stems directly from the deterministic PDE solver (the FEM error in our case) \cite{Cliffe2011,TeckentrupMLMC2013} and as such it can be estimated empirically, i.e.~by experimenting with the mesh size and time step, and/or by using theoretical error estimates \cite{brenner2007mathematical}. Here, we estimate the bias empirically (cf.~description of the MLMC algorithm) and we check that the empirical bias convergence order with $h$ and $\Delta t$ is second-order as expected given our discretization.
\end{remark}

For MLMC, we first need to estimate the optimal number of samples on each level for a given tolerance $\varepsilon$. Let $C_\ell$, $V_\ell$ be the cost and variance of one sample $Q_\ell(\cdot) - Q_{\ell-1}(\cdot)$ respectively. The total MLMC cost and variance are
\begin{align}
\label{eq:MLMC_ctot_vartot}
C_{tot} = \sum\limits_{\ell = 1}^LN_\ell C_\ell,\quad \hat{V} = \sum\limits_{\ell = 1}^LN_\ell^{-1}V_\ell.
\end{align}
It is possible to minimize the estimator variance for a fixed total cost \cite{giles2015multilevel}. For a fixed MSE tolerance $\varepsilon^2$, the optimal number of samples for each level and related total cost are,
\begin{align}
\label{eq:optimal_number_of_samples}
N_\ell = \left((1-\theta)^{-1}\varepsilon^{-2}\sum\limits_{l = 1}^L\sqrt{V_l C_l}\right)\sqrt{V_\ell/C_\ell},\quad
C_{tot} = (1-\theta)^{-1}\varepsilon^{-2}\left(\sum\limits_{\ell = 1}^L\sqrt{V_\ell C_\ell}\right)^2.
\end{align}

We can now describe the MLMC algorithm used in this work:

\paragraph{MLMC algorithm (taken from \cite{giles2015multilevel})}
\begin{itemize}
	\item Set the required tolerance $\varepsilon$, $\theta\in(0,1)$, the maximum level $L_{\max}$, the initial number of levels $L$ and the initial number of samples $\bar{N}_\ell$ to be taken on each.
	\item \textbf{while} extra samples need to be evaluated ($\exists \ell:\ \bar{N}_\ell > 0$):
	\begin{enumerate}
		\item for each level, evaluate all the samples scheduled to be taken;
		\item compute/update estimates for the level variance $V_\ell$, $\ell=1,\dots,L$;
		\item compute optimal number of samples $N_\ell$ by using \eqref{eq:optimal_number_of_samples} and update the numbers of extra samples needed $\bar{N}_\ell$ accordingly;
		\item check whether the weak error (i.e.~the bias) ${|\E[\hat{P} - P]|}$ is below the required tolerance $\sqrt{\theta}\varepsilon$; to see how the bias is estimated in practice without knowing the exact solution, see Section 3 in \cite{giles2015multilevel}.
		\item if not: if $L=L_{\max}$ report failed convergence; otherwise set $L := L + 1$, update $N_\ell$ and $\bar{N}_\ell$ and compute $N_L=\bar{N}_L$ (again using \eqref{eq:optimal_number_of_samples}).
	\end{enumerate}
\end{itemize}

\begin{remark}
\label{rem:comparing_efficiency}
The standard way \cite{GilesWaterhouse2009,giles2015multilevel} to compare the efficiency of these algorithms is to fix the same tolerance for both methods and compare their total costs, given by $C_{\text{tot}}=\bar{C}NM$ for QMC and by $C_{tot} = \sum_{\ell = 1}^LN_\ell C_\ell$ for MLMC (cf.~\eqref{eq:MLMC_ctot_vartot}). Furthermore, to ensure that the bias level is the same for both methods, the QMC routine must be run on the finest mesh and time step size of the MLMC hierarchy. This gives $\bar{C}=C_L$. In practice, for the sake of comparing methods, the actual costs $C_\ell$, $\ell=1,\dots,L$ can be replaced by pseudo-costs, i.e.~by setting $C_\ell\approx \hat{c}_3 2^{\hat{\gamma}\ell}$, where $\hat{c}_3$ and $\hat{\gamma}$ are the values, estimated with CPU timings, of the constants $c_3$ and $\gamma$ appearing in Theorem \ref{th:MLMC}. We use this latter approach.
\end{remark}

\section{Numerical results}
\label{sec:num_brain_results}

We now compare the efficiency of standard MC, QMC and MLMC when employed to solve Models 1 and 2. In what follows, we let $\mathcal{T} = \{30k \text{ min}\}_{k=1}^{k=48}$ and define
\begin{align}
\mathcal{Q}=\{Q_g(t),\ t\in \mathcal{T}\} \cup \{Q_w(t),\ t\in \mathcal{T}\} \cup \{q_g(t),\ t\in \mathcal{T}\} \cup \{q_w(t),\ t\in \mathcal{T}\},
\end{align}
to be the set of all the output functionals of interest considered (cf~\eqref{eq:Q_gw} and \eqref{eq:q_gw}).

\subsection{Estimation of MLMC parameters}

We first estimate the MLMC parameters $\alpha$, $\beta$ and $\gamma$ of Theorem \ref{th:MLMC}. Since we are considering the estimation of multiple output functionals, we estimate $\alpha$ and $\beta$ by monitoring the bias and variance at each level $\ell$:
\begin{align}
\label{eq:comparison_error_measures}
\max\limits_{Q\in\mathcal{Q}}|\E[Q_\ell - Q_{\ell-1}]|,\quad \max\limits_{Q\in\mathcal{Q}}\V[Q_\ell - Q_{\ell-1}].
\end{align}
We expect $\alpha=2$, $\beta=2\alpha=4$ and $\gamma=4$ since for random PDEs we have that the bias convergence is typically the same as the deterministic FEM convergence order, $\beta=2\alpha$ \cite{Cliffe2011,TeckentrupMLMC2013}, the numerical method is second-order in both time and space and we are using a multigrid-preconditioned Krylov method (cf.~Section \ref{sec:numerical_details_stochastic_models}). The value for $\gamma$ stems from the fact that the number of time steps on level $\ell$ are proportional to $2^\ell$ and the linear solver used has (essentially) linear complexity in the number of degrees of freedom, which in turn scale proportionally to $2^{3\ell}$. We therefore get a cost per level proportional to $2^{(3 + 1)\ell}$ and a $\gamma=4$.

To estimate the bias and variance in \eqref{eq:comparison_error_measures} we take $N=4000$ samples on the first two levels and $N=100$ on the finest level. The choice of the number of samples on the finest level is motivated by the following considerations. The computational resources available were limited, cf.~Section \ref{sec:solver_software_limitedCPU}. The number of vertices on the finest level of the MLMC hierarchy is large (cf.~Table \ref{tab:hierarchy_meshes}), resulting in a memory burden of $\approx 50$ GB to load the mesh, the box mesh in which the brain mesh is embedded (cf.~Section~\ref{sec:background}) and the FEM subspaces. Additionally, solving one instance of \eqref{eq:transport} on this mesh takes more than $24$ hours in serial. The overall sampling procedure took around $2$ weeks of combined use of the Oxford Mathematical Institute servers and the University of Oslo Abel cluster. The same computation performed in serial would have taken roughly two years. Adding a finer level to the hierarchy would have therefore been prohibitive.

\begin{figure}[h!]
	\centering
	\includegraphics[width=\textwidth]{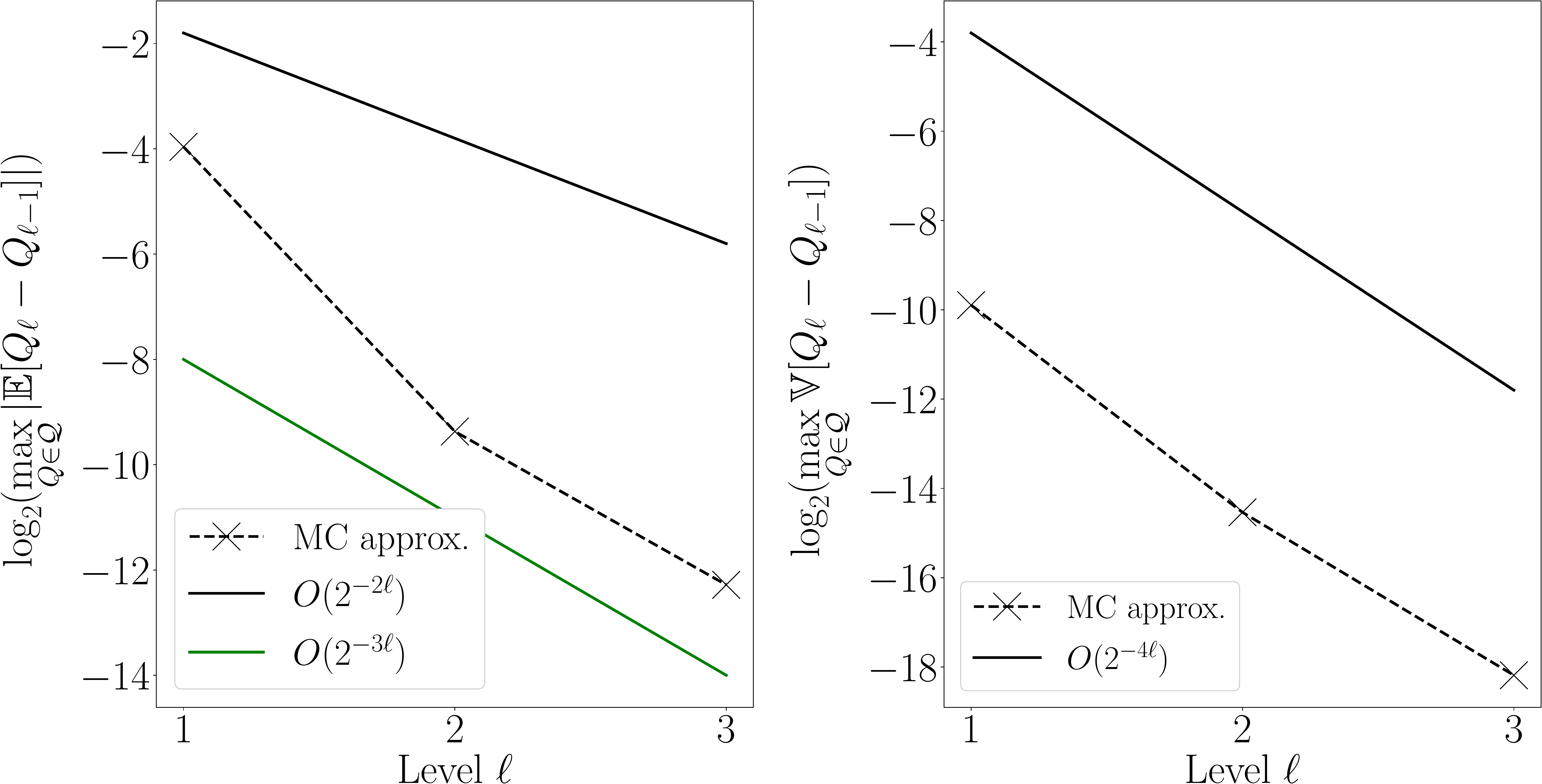}
	\centering
	\caption{\textit{Convergence behaviour of the FEM approximation to the solution of Model 1. The estimated convergence order for the variance agrees with our predictions and with what expected by the theory in the diffusion-only case \cite{TeckentrupMLMC2013}. The bias convergence order observed is instead higher than expected. Estimated parameters via linear regression: $\hat{\alpha}\approx 4.16$, $\hat{\beta}\approx 4.15$.}}
	\label{fig:comparison1}
\end{figure}

\begin{figure}[h!]
	\centering
	\includegraphics[width=\textwidth]{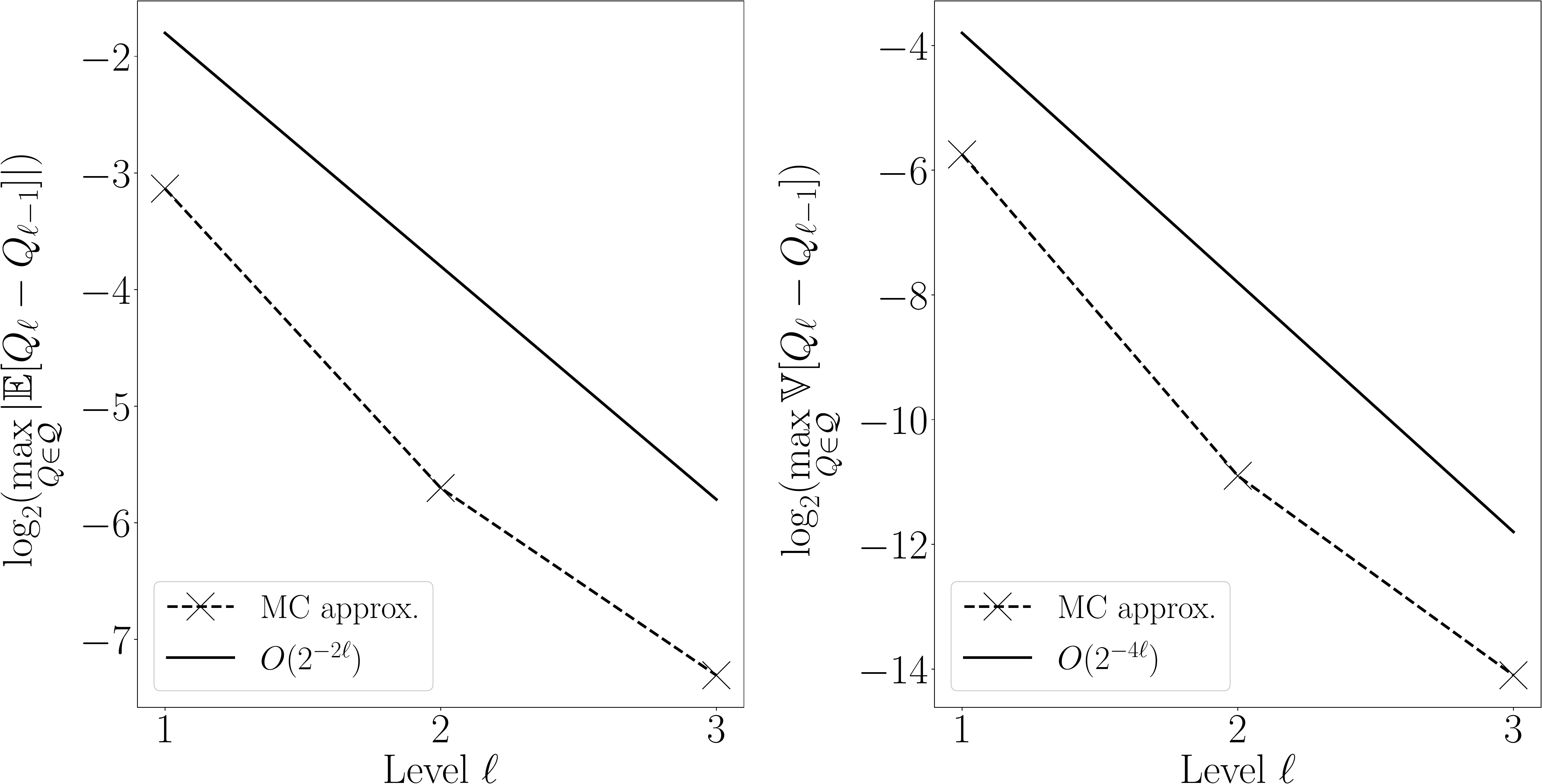}
	\centering
	\caption{\textit{Convergence behaviour of the FEM approximation to the solution of Model 2. The estimated convergence orders agree with our predictions and with what expected by the theory in the diffusion-only case \cite{TeckentrupMLMC2013}. Estimated parameters via linear regression: $\hat{\alpha}\approx 2.09$, $\hat{\beta}\approx 4.18$.}}
	\label{fig:comparison2}
\end{figure}

Figures \ref{fig:comparison1} and \ref{fig:comparison2} show the bias
and variance versus refinement level for Models 1 and 2,
respectively. We observe that the numerical estimates closely match
the theoretical expectations. The estimated variance convergence order
for both models is $\hat{\beta} \approx 4.2$, which is just above the
theoretical value of $\beta = 4$. For Model 2, the estimated bias
order is $\hat{\alpha} = 2.09$ which again closely matches the
theoretical estimate of $\alpha = 2$. However for Model 1, we observe
that the bias decays more rapidly than expected ($\hat{\alpha} = 4.16$
versus $\alpha = 2$). In this case, we are likely observing a
pre-asymptotic regime and the convergence rate seems to be decaying as
$\ell$ increases; this notion would be in agreement with theoretical
convergence
results~\cite{brenner2007mathematical,TeckentrupMLMC2013}. Finally,
when estimating $\gamma$ as taking the average wall-clock time of each
sample as a proxy (data not shown), we obtain $\hat{\gamma} \approx
4.09$, which is close to the theoretical prediction of $\gamma=4$.

\subsection{Mean square error weighting under limited computational resources}
\label{sec:mean_sq_err_weighting_limited}

Note that we have a finite number of meshes available and consequently the version of MLMC considered here is ``weaker'' than the true MLMC algorithm, described in Section \ref{sec:QMC_MLMC_algorithms}, since the maximum level $L$ is bounded, cf.~Remark \ref{rem:max_level_bounded}. In fact, we are unable to reduce the bias below the threshold given by the finest mesh of the hierarchy without resorting to more advanced techniques (see later Remark \ref{rem:Richardson-Romberg}). However, we can still balance the relative weight of bias and statistical error by choosing two different values of the MLMC weight parameter $\theta$ (cf.~\cite{haji2016optimization} and Section \ref{sec:background}). Based on the fact that for this problem: 1) the maximum $L$ is restricted; 2) $N_L$ is restricted; 3) we have estimated values for the MLMC parameters $\alpha$, $\beta$, and $\gamma$ (cf.~Theorem \ref{th:MLMC}); we can estimate \emph{a priori} the largest possible values of $\theta$ that we can use without making the number of samples on the finest level exceed $100$. This yields the values $\theta = 0.041$ for Model 1 and $\theta = 0.72$ for 2, corresponding to the best achievable MSE tolerances of roughly $\varepsilon=3.3\times 10^{-4}$ for Model 1 and of $\varepsilon=2.5\times 10^{-3}$ for Model 2. Note that in the Model 1 case, the bias is much smaller (compare figures \ref{fig:comparison1} and \ref{fig:comparison2}), hence why the chosen $\theta$ is smaller as well.

\begin{remark}
	Note that the bias (i.e.~the discretization error) can be estimated by the MLMC algorithm without requiring a reference or exact solution (recall the MLMC algorithm in Section \ref{sec:QMC_MLMC_algorithms}) and is the same independently from which Monte Carlo method is used. Furthermore, all methods can compute their own estimates of the statistical error. For this reason, no reference or exact solution is needed to compute the MSE or to construct the figures shown in this section.
\end{remark}

\begin{remark}
	\label{rem:Richardson-Romberg}
	In practice, it is possible to reduce the MLMC estimator bias by augmenting MLMC with Richardson-Romberg extrapolation \cite{giles2008,lemaire2017multilevel}. However, we leave this enhancement for future research.
\end{remark}

\subsection{Tracer evolution over time}

\begin{figure}[!h]
	\begin{center}
		\includegraphics[width=\linewidth]{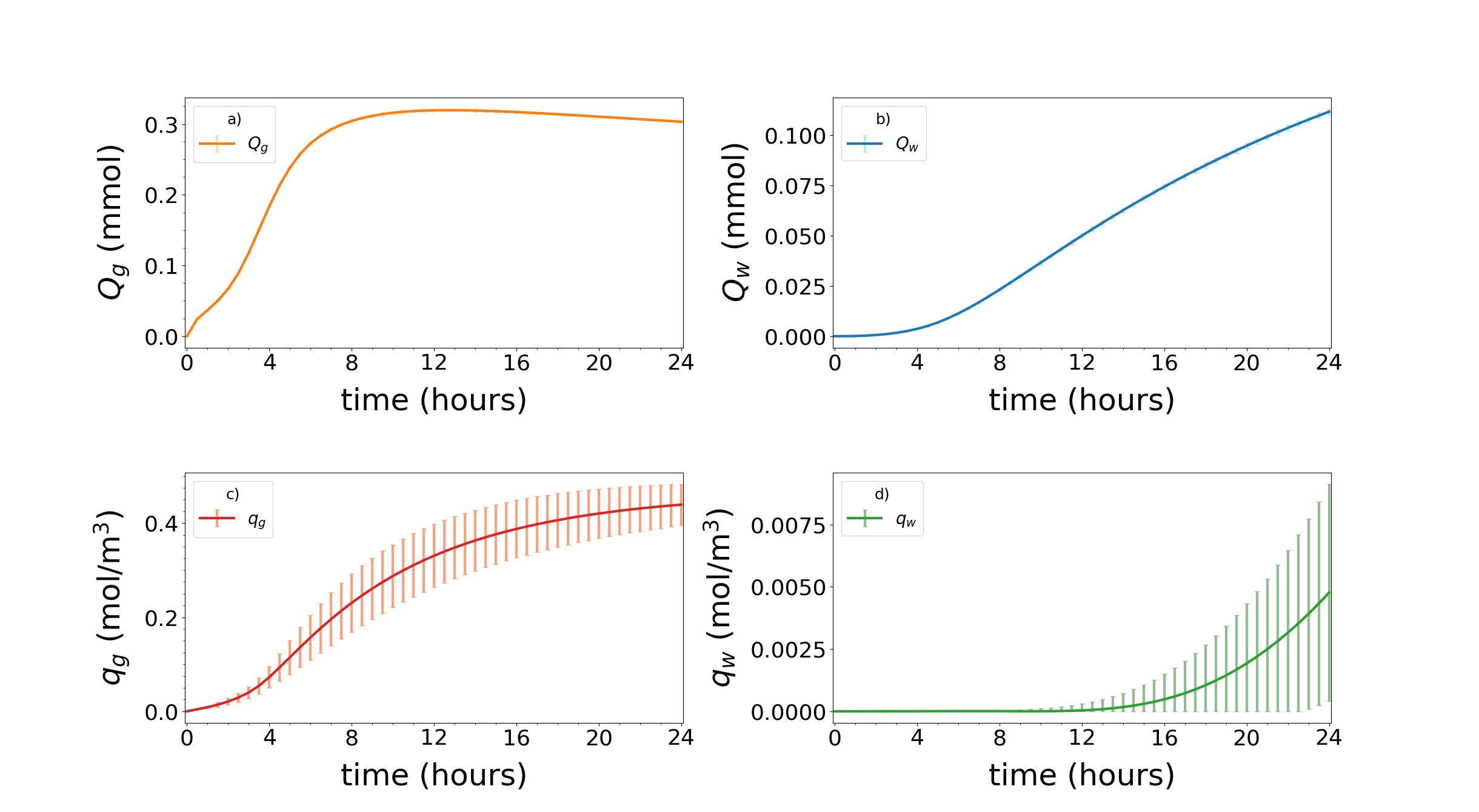}
		\caption{\textit{Mean and standard deviation vs time for all quantities of interest for Model 1. The continuous lines correspond to expected values and the vertical bars indicate plus or minus one standard deviation away from the mean (negative values are excluded here). Both were computed using MLMC with a MSE tolerance of $\varepsilon=3.3\times 10^{-4}$, which is the best achievable tolerance given the discretization error and the restrictions on the number of samples on the finest level, cf.~Section \ref{sec:mean_sq_err_weighting_limited}.}}
		\label{fig:mlmcavg_model1}
	\end{center}
\end{figure}

\begin{figure}[!h]
	\begin{center}
		\includegraphics[width=\linewidth]{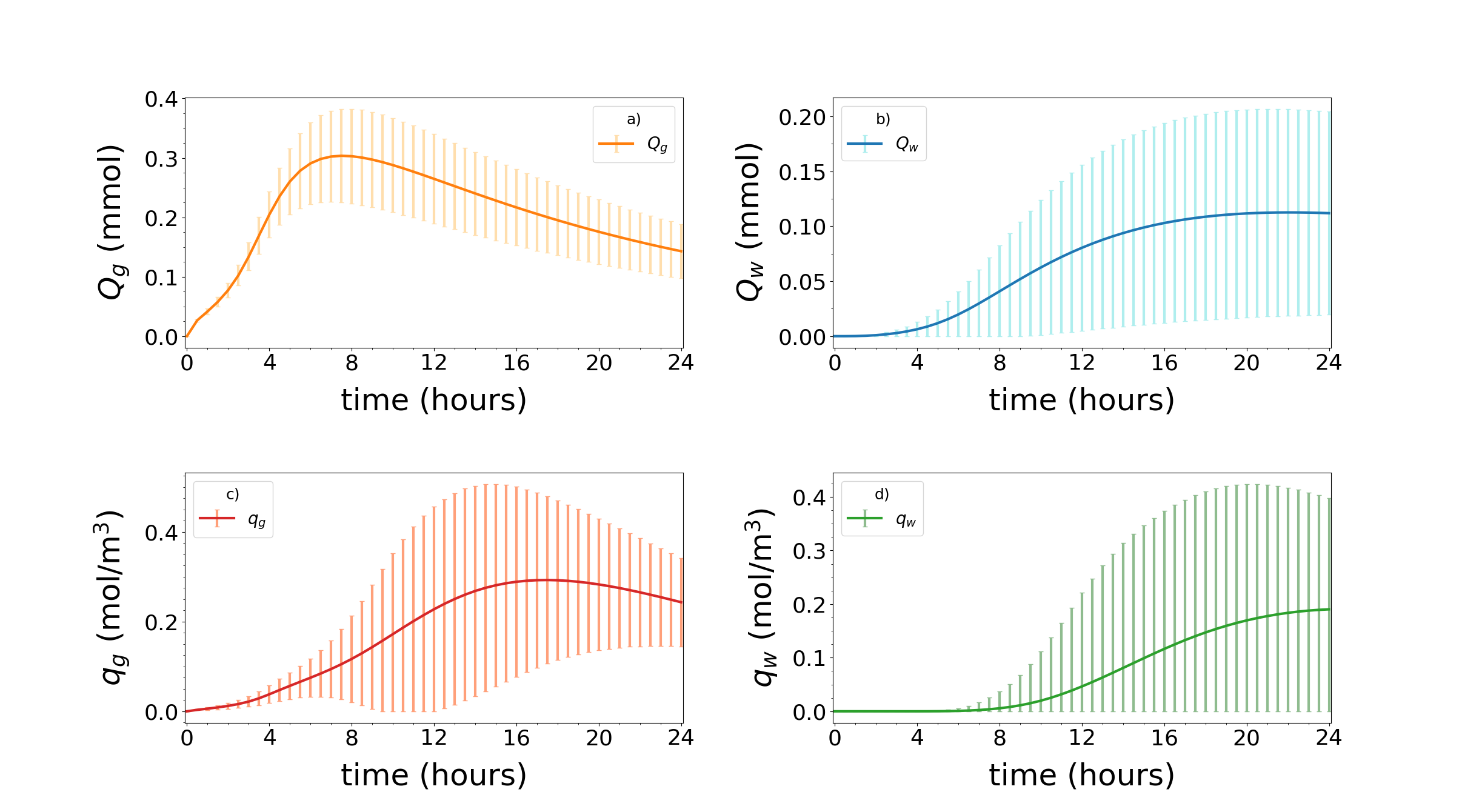}
		\caption{\textit{Mean and standard deviation vs time for all quantities of interest for Model 2. The continuous lines correspond to expected values and the vertical bars indicate plus or minus one standard deviation away from the mean (negative values are excluded here). Both were computed using MLMC with a MSE tolerance of $\varepsilon=2.5\times 10^{-3}$, which is the best achievable tolerance given the discretization error and the restrictions on the number of samples on the finest level, cf.~Section \ref{sec:mean_sq_err_weighting_limited}.}}
		\label{fig:mlmcavg_model2}
	\end{center}
\end{figure}

In Figures \ref{fig:mlmcavg_model1} and \ref{fig:mlmcavg_model2} we
present the expected value and standard deviation of all quantities of
interest as a function of time for both models. These results have
been obtained with MLMC by setting the lowest achievable MSE
tolerances given the restriction on the maximum number of samples on
the finest level, cf.~Section
\ref{sec:mean_sq_err_weighting_limited}. The total number of samples
used for these simulations are $N_\ell=[26060,\ 1372,\ 100]$ for Model
1 and $N_\ell=[26567,\ 1171,\ 100]$ for Model 2 for $\ell = 1, 2, 3$.
We remark that the figure errorbars represent one standard deviation
away from the mean, and not necessarily the actual variability of
these quantities.

In both models, tracer first spreads to the gray matter, reaching a
peak amount of tracer in 8--12 hours (Figures
\ref{fig:mlmcavg_model1}a, \ref{fig:mlmcavg_model2}a). Over time, the
tracer also spreads into the white matter (Figures
\ref{fig:mlmcavg_model1}b, \ref{fig:mlmcavg_model2}b). The
time-to-peak amount of tracer in white matter is estimated at
$\approx$ 24 hours in Model 2 (Figure \ref{fig:mlmcavg_model2}b),
while for Model 1 this peak would occur considerably later (Figure
\ref{fig:mlmcavg_model1}b). In Figure \ref{fig:mlmcavg_model1}, we
observe that for Model 1, the variances of the global amounts of
tracer $Q_g$ and $Q_w$ are small. However, the variability in the
local tracer concentration $q_g$ and $q_w$ is much larger and, as
such, substantially contributes to the computational complexity of the
Monte Carlo simulations. We observe that for Model 1, the local
variability of the random coefficients "averages out" when computing
global output functionals. Thus, if global quantities were the only
focus, a MC simulation with small sample size could be sufficient to
achieve good accuracy for this model. Conversely, we note that all
quantities of interest for Model 2 (amount of tracer in gray and white
matter, and tracer concentrations in local regions) have large
standard deviations (Figure~\ref{fig:mlmcavg_model2}).

Comparing these results with the MSE tolerances imposed we can
estimate the number of correct digits of the computed expectations:
the estimates obtained for Model 1 are roughly within $3$ digits of
accuracy for the gray matter quantities $Q_g$ and $q_g$, $2$ digits
for $Q_w$ and $1$ digit only for $q_w$. Similarly, all Model 2 gray
matter quantities $Q_g$ and $q_g$ are roughly correct within $2$
digits, while the white matter outputs $Q_w$ and $q_w$ are just
slightly less accurate.

%

\subsection{Comparison of MC, QMC and MLMC performance}

\begin{figure}[h!]
	\centering
	\includegraphics[width=\textwidth]{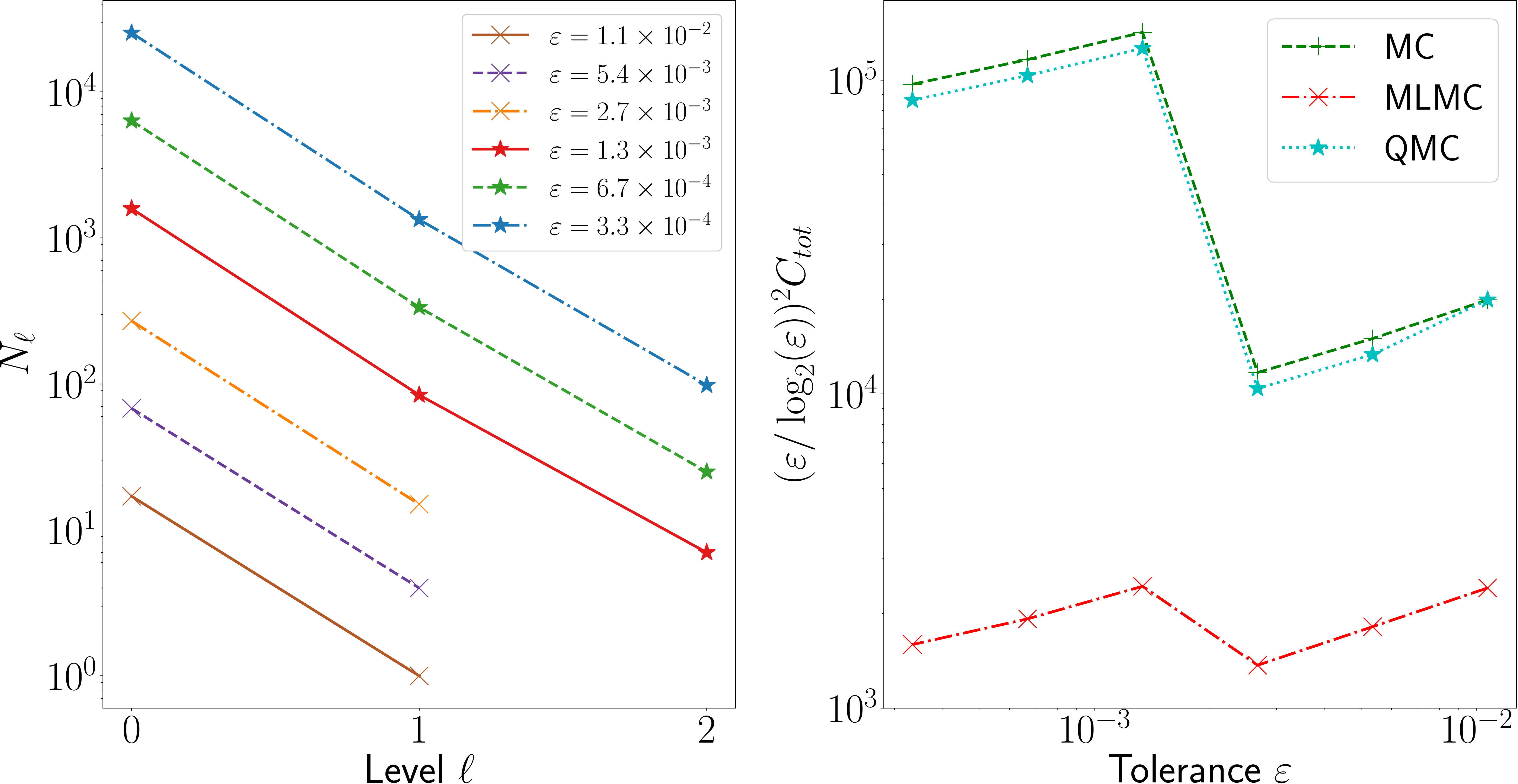}
	\centering
	\caption{\textit{Convergence of standard MC, QMC and MLMC for the solution of Model 1 ($\theta=0.041$). In the plot on the left we show how the MLMC algorithm automatically selects the optimal number of samples $N_{\ell}$ on each level to achieve a given tolerance $\varepsilon$. In the plot on the right we compare the efficiency of the methods for different tolerances. The savings of MLMC with respect to standard MC and QMC are considerable, while QMC barely improves on standard MC (see main text).}}
	\label{fig:comparison3}
\end{figure}

\begin{figure}[h!]
	\centering
	\includegraphics[width=\textwidth]{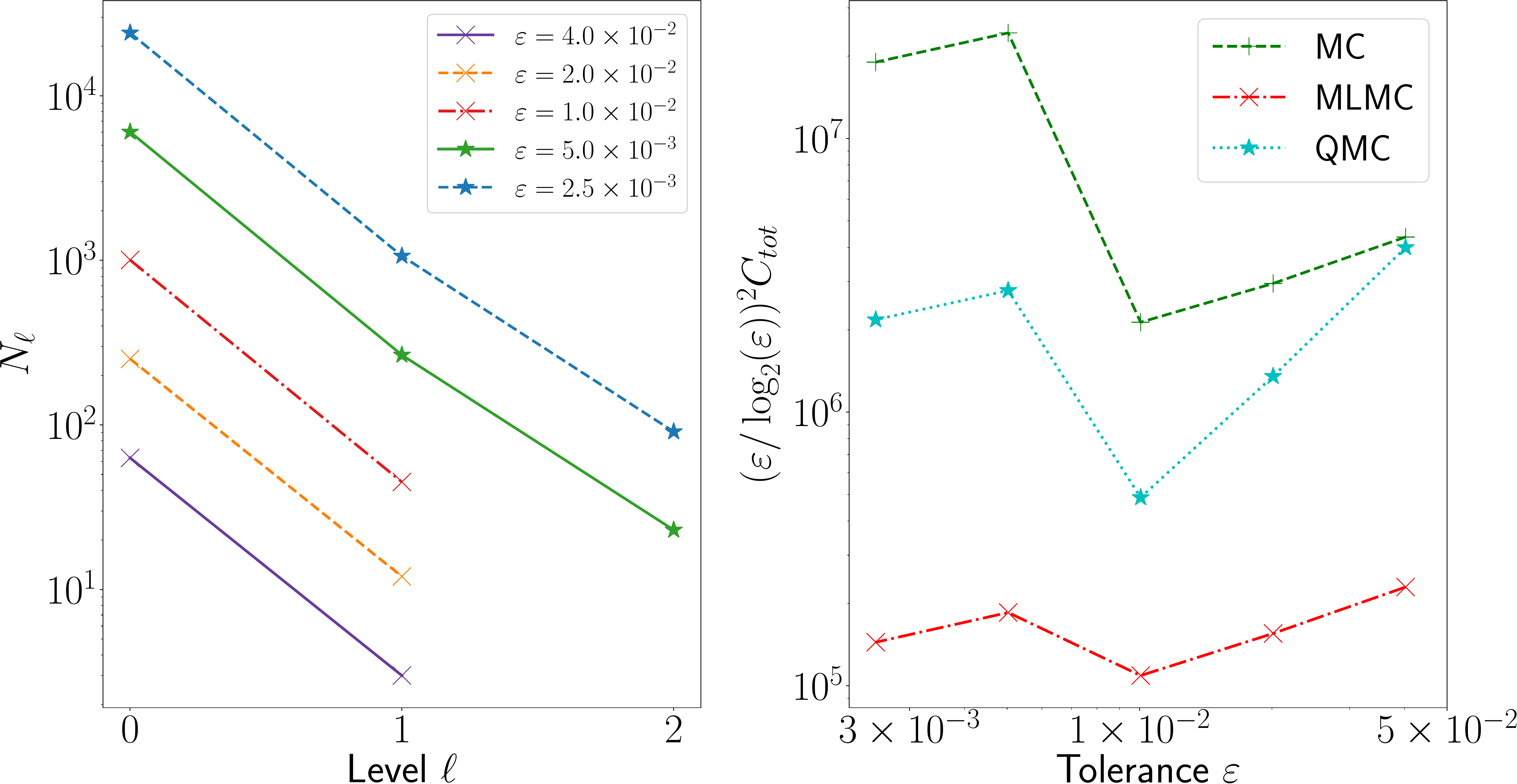}
	\centering
	\caption{\textit{Convergence of standard MC, QMC and MLMC for the solution of Model 2 ($\theta=0.72$). In the plot on the left we show how the MLMC algorithm automatically selects the optimal number of samples $N_{\ell}$ on each level to achieve a given tolerance $\varepsilon$. In the plot on the right we compare the efficiency of the methods for different tolerances. MLMC significantly outperforms QMC, which in turn considerably outperforms standard MC.}}
	\label{fig:comparison4}
\end{figure}

In Figures \ref{fig:comparison3} and \ref{fig:comparison4} (left) we show the optimal number of samples chosen automatically by MLMC on each level as the root mean square error tolerance $\varepsilon$ is reduced. The maximum level chosen is increased as $\varepsilon$ decreases in order to satisfy the bias tolerance. Note that the smallest values of $\varepsilon$ considered correspond to the lowest bias tolerance that standard MLMC can achieve with an upper limit of $100$ samples on the finest level (cf.~Remark \ref{rem:Richardson-Romberg}).

In Figures \ref{fig:comparison3} and \ref{fig:comparison4} (right), we compare the total computational cost $C_{tot}$ of standard MC, MLMC and QMC for the solution of Model 1 and 2, respectively. The cost of running a full standard MC simulation for this problem is prohibitive (on the order of $100$ years in serial!) so we estimate the MC cost by using the level variance and cost. This is a standard procedure and essentially derives from the application of the central limit theorem \cite{giles2015multilevel}. We find that QMC is significantly slower than MLMC. For this reason we only estimate the number of QMC samples needed by running QMC on the second finest mesh of the hierarchy rather than on the finest (cf.~Remark \ref{rem:comparing_efficiency}). The difference should be minimal since the number of samples needed appears to be approximately constant on the finer levels of the hierarchy in numerical experiments \cite{CrociMLQMC}. Since $\beta=\gamma$, Theorem \ref{th:MLMC} predicts an overall MLMC complexity of $\varepsilon^{-2}(\log \varepsilon)^2$ for a root mean square error tolerance $\varepsilon$. We therefore expect a near constant cost curve for $\varepsilon^{2}(\log \varepsilon)^{-2}C_{tot}$ versus $\varepsilon$ in the MLMC case. The numerical results corroborate the theoretical expectations: while the MLMC cost lines oscillate some, they are well-fitted by a horizontal line (estimated slope $\approx 0.05$ for Model 1 and $\approx 0.1$ for Model 2). Overall, for both models MLMC significantly outperforms both QMC and standard MC, with a $O(100)$ factor of improvement with respect to standard MC.

While the qualitative behaviour of standard MC and MLMC is consistent between the two models, QMC behaves differently. For Model 2 (Figure \ref{fig:comparison4}), the performance of QMC considerably improves on that of standard MC, especially for smaller MSE tolerances. On the other hand, for Model 1 (Figure \ref{fig:comparison3}), the improvement is negligible and QMC performs essentially the same as standard MC.

This behaviour could be interpreted in the context of the formulations of Models 1 and 2 (cf.~Section \ref{sec:coefficient_models}). While the stochastic input in Model 2 includes $1$ random field and $1$ random variable, Model 1 depends on $4$ random fields and $1$ random variable. Given the lack of performance gain for QMC applied to Model 1, we hypothesize that the higher input dimensionality affects the QMC convergence, causing the rate to decay to a standard $O(N^{-1/2})$ MC rate. Indeed, the fact that QMC performance degrades with high input dimensions is well-known \cite{CaflishMorokoffOwen}. It therefore appears that the (ML)QMC method presented in \cite{CrociMLQMC} is not robust with respect to the number of random field inputs, at least in 3D where the dimensionality is larger. We did not observe this ill-behaviour in analogous numerical tests performed on a convection-diffusion PDE with random coefficients on a square domain.

\begin{remark}
  Adding a coarser level to the mesh hierarchy, given by the original
  Colin27 human adult brain atlas mesh \cite{FangEtAl2010} (version 2)
  did not improve the performance of MLMC. This was determined by first estimating the cost and variance of the coarsest MLMC levels in the extended hierarchy using $4000$ MC pilot samples and then determining which hierarchy would give the best savings in terms of cost (see e.g.~Section 2.6 in~\cite{giles2015multilevel}).
\end{remark}

\section{Discussion}
\label{sec:discussion}

Most theoretical UQ research on equations with random coefficients focuses on the same model problem~\cite{TeckentrupMLMC2013,CharrierMLMC2013,KuoSchwabSloan2015,KuoScheichlSchwabEtAl2017}, with numerical results often performed on simple boxes and in low dimensions. For this reason, and because of their different requirements and convergence properties, it is a priori unknown which method performs best for a given problem. While asymptotic complexity results might be available \cite{giles2015multilevel,TeckentrupMLMC2013,KuoSchwabSloan2015}, it might not be possible to observe the asymptotic behaviour due to practical limitations that arise in large-scale problems (e.g.~limited computational resources, small hierarchy size, poor low-fidelity models). In fact, when considering large-scale applications with physics-inspired problems on MRI-derived geometries, currently, the only way to compare the performance of different methods is via numerical experiments. Nevertheless, the application of UQ to realistic large-scale problems is scarce in the literature, to the extent that (that we know of) this is the first study employing advanced UQ techniques to study the human brain.

The only related study that we are aware of comes from our previous work on the subject~\cite{CrociVinjeRognes2019}. In this study we found that diffusion alone could not explain MRI data, suggesting flow velocities may impact the distribution of tracer seen in MRI data, possibly also affecting both brain clearance and drug delivery. However, in~\cite{CrociVinjeRognes2019} standard MC was used, which forced us to resort to coarser meshes and shorter simulation time. The need for a more efficient UQ methodology motivated the current study.

In terms of limitations, in this study we considered a single brain geometry deriving from the Colin27 mesh \cite{colin27}. A set of different geometries should be considered to ensure that the results obtained are patient-independent. We only modelled a single fluid flow compartment, while future research could consider more elaborate models accounting for all the different fluids flowing in the parenchyma (venous, arterial and capillary blood, interstitial fluid, cerebrospinal fluid). The computational resources available for this study were limited and with more computational power we could have obtained more accurate results by adding a finer level to the hierarchy and/or increasing the number of samples. However, we did not encounter any significant issues with setting up a hierarchy on which good MLMC convergence could be observed, a problem mentioned in recent work by Quaglino et al.~\cite{Quaglino2018}, in which they apply MFMC for a cardiac electrophysiology application.

We also point at a limitation related to the choice of covariance
structure for the diffusion and velocity fields. Here, these fields
have been assigned an isotropic covariance. As such, the correlation
between any two points only depends on their Euclidean distance, and
not on the (domain) interior path length between the two
points. Consequently, brain surface regions such as the different gyri
that are closer to each other than the field correlation length,
become correlated even if separated by a sulcus. Since the average
sulcal width is estimated to range between $1.5$ and $3$ mm in
humans~\cite{JinEtAl2018,Madan2019}, this aspect does not play a role
for the velocity field ($\lambda\approx 1$mm), but for the diffusion
coefficient it may ($\lambda= 1$cm). We expect global quantities
(e.g.~$Q_g$ and $Q_w$) and local quantities depending on white matter
regions which are far from the brain surface (e.g.~$q_w$) to be
largely unaffected by this modeling choice. Conversely, local
quantities that depend on gray matter local regions (e.g.~$q_g$) might
be affected and further investigation would be needed. We remark,
however, that isotropic covariances make for very efficient sampling
methods, and designing and efficiently sampling a random field which is
topology-aware at the high resolution scale of the sulci is highly
challenging.

Overall, the random fields parameters were chosen to represent the current understanding of the relevant physiology to the extent possible, with physical considerations as the primary factor in the choice of fields and parameters. While most of the model parameters do not affect the performance of the numerical solver, QMC or MLMC in any way, the following aspects should be considered for  alternative models:
\begin{compactitem}
\item
  \textbf{Velocity magnitude and diffusion coefficient:} a much larger velocity magnitude (or a much smaller diffusion coefficient) would increase the P\'eclet number and harm the stability of the finite element solver used (cf. Section \ref{sec:Peclet}). A more advanced numerical solver would be needed in this case. We remark that for random PDEs the convergence of MLMC directly depends on the convergence properties of the deterministic solver used \cite{Cliffe2011,TeckentrupMLMC2013}. 
\item
  \textbf{Random field variance:} larger variability implies that more samples would be needed for an accurate result. This does not affect the efficacy of the methods, but it does increase the CPU hours needed for the computation, which might be a problem when the computational resources are limited.
\item
  \textbf{Smoothness parameter and correlation length:} small values of these parameters are known to affect most random field sampling techniques and extra care must be taken. At the same time small values would also introduce sharp and abrupt spatial changes in the sampled field that are acceptable for e.g.~oil reservoir modelling \cite{potsepaev2010application}, but would be nonphysical in a healthy brain. We remark that the velocity correlation length chosen in this study is quite small, showing that the sampling techniques in \cite{Croci2018,CrociMLQMC} still work well in this case.
\end{compactitem}

\section{Conclusions}
\label{sec:conclusions}

We have compared and evaluated the MC, QMC, and MLMC uncertainty
quantification and sampling methods for two physiologically realistic
brain transport models. Even under restriction on the maximum number
of samples on the finest mesh level and on the finest mesh available,
MLMC significantly outperforms all other methods, yielding an
improvement factor of roughly $100$ with respect to standard MC. QMC
outperforms standard MC by a factor of approximately $10$ for one of
the models (Model 2), but yields no performance gains for the other
(Model 1). Overall, for this application, MLMC achieves the best
performance and should be preferred. We remark that a standard MC
simulation would have taken on the order of $10$ years to complete in
parallel ($O(100)$ years in serial!), while only $2$ weeks are needed
for MLMC.

MLQMC methods were not investigated directly in this study. However,
our numerical findings indicate that no additional improvement can be
achieved in the Model 1 case. On the other hand, Model 2 seems
amenable to MLQMC. Thus, MLQMC could also bring additional
computational gains for similar models with a small number of input
random fields.

In conclusion, even large-scale problems on MRI-derived geometries are amenable to advanced Monte Carlo methods and we hope that our work can inspire further research on uncertainty quantification for biomedical engineering problems.

\section*{Acknowledgements}
This research is supported by the European Research Council via Grant \#714892 (Waterscales), the Research Council of Norway via Grant \#250731 (Waterscape), by the NOTUR grant NN9316K, and by the EPSRC Centre For Doctoral Training in Industrially Focused Mathematical Modelling (EP/L015803/1). Part of this work was performed on the Abel Cluster, owned by the University of Oslo and Uninett/Sigma2, and operated by the Department for Research Computing at USIT, the University of Oslo IT-department \url{http://www.hpc.uio.no/}.

\printbibliography

@article{JinEtAl2018,
		author = {Jin, Kaide and Zhang, Tianqi and Shaw, Marnie and Sachdev, Perminder and Cherbuin, Nicolas},
		doi = {10.3389/fnagi.2018.00339},
		issn = {1663-4365},
		journal = {Frontiers in Aging Neuroscience},
		month = {nov},
		pages = {339},
		publisher = {Frontiers Media SA},
		title = {{Relationship between sulcal characteristics and brain aging}},
		volume = {10},
		year = {2018}
	}

@article{Madan2019,
		author = {Madan, Christopher R.},
		doi = {10.1186/s40708-019-0098-1},
		issn = {21984026},
		journal = {Brain Informatics},
		month = {dec},
		number = {1},
		pages = {1--11},
		publisher = {Springer Berlin Heidelberg},
		title = {{Robust estimation of sulcal morphology}},
		url = {https://doi.org/10.1186/s40708-019-0098-1},
		volume = {6},
		year = {2019}
	}

@article{sokoloff1989circulation,
		author =        {Sokoloff, Lois},
		journal =       {Basic neurochemistry},
		pages =         {338--413},
		publisher =     {Raven Press New York},
		title =         {Circulation and energy metabolism of the brain},
		volume =        {2},
		year =          {1989},
	}

@article{IliffEtAl2012,
		author =        {Iliff, Jeffrey J and Wang, Minghuan and
			Liao, Yonghong and Plogg, Benjamin A and Peng, Weiguo and
			Gundersen, Georg A and Others},
		journal =       {Science Translational Medicine},
		number =        {147},
		pages =         {147ra111},
		title =         {{A paravascular pathway facilitates {CSF} flow
				through the brain parenchyma and the clearance of
				interstitial solutes, including amyloid beta}},
		volume =        {4},
		year =          {2012},
	}

@article{JessenEtAl2015,
		author =        {Jessen, Nadia Aalling and Munk, Anne Sofie Finmann and
			Lundgaard, Iben and Nedergaard, Maiken},
		journal =       {Neurochemical Research},
		number =        {12},
		pages =         {2583--2599},
		publisher =     {Springer},
		title =         {{The glymphatic system: a beginner's guide}},
		volume =        {40},
		year =          {2015},
	}

@article{IliffEtAl2013,
		author =        {Iliff, Jeffrey J and Wang, Minghuan and
			Zeppenfeld, Douglas M and Venkataraman, Arun and
			Plog, Benjamin A and Liao, Yonghong and Deane, Rashid and
			Nedergaard, Maiken},
		journal =       {Journal of Neuroscience},
		number =        {46},
		pages =         {18190--18199},
		publisher =     {Soc Neuroscience},
		title =         {{Cerebral arterial pulsation drives paravascular
				CSF--interstitial fluid exchange in the murine
				brain}},
		volume =        {33},
		year =          {2013},
		isbn =          {1529-2401 (Electronic) 0270-6474 (Linking)},
	}

@article{CarareEtAl2008,
		author =        {Carare, R O and Bernardes-Silva, M and Newman, T A and
			Page, A M and Nicoll, J A R and Perry, V H and
			Weller, R O},
		journal =       {Neuropathology and Applied Neurobiology},
		number =        {2},
		pages =         {131--144},
		publisher =     {Wiley Online Library},
		title =         {{Solutes, but not cells, drain from the brain
				parenchyma along basement membranes of capillaries
				and arteries: significance for cerebral amyloid
				angiopathy and neuroimmunology}},
		volume =        {34},
		year =          {2008},
	}

@article{aldea2019cerebrovascular,
		author =        {Aldea, Roxana and Weller, Roy O and Wilcock, Donna M and
			Carare, Roxana O and Richardson, Giles},
		journal =       {Frontiers in Aging Neuroscience},
		publisher =     {Frontiers Media SA},
		title =         {{Cerebrovascular smooth muscle cells as the drivers
				of intramural periarterial drainage of the brain}},
		volume =        {11},
		year =          {2019},
	}

@article{SmithEtAl2017,
		author =        {Smith, Alex J and Yao, Xiaoming and Dix, James A and
			Jin, Byung-Ju and Verkman, Alan S},
		journal =       {eLife},
		number =        {e27679},
		pages =         {16},
		publisher =     {eLife Sciences Publications Limited},
		title =         {{Test of the `glymphatic' hypothesis demonstrates
				diffusive and aquaporin-4-independent solute
				transport in rodent brain parenchyma}},
		volume =        {6},
		year =          {2017},
	}

@article{wolak2013diffusion,
		author =        {Wolak, Daniel J and Thorne, Robert G},
		journal =       {Molecular pharmaceutics},
		number =        {5},
		pages =         {1492--1504},
		publisher =     {ACS Publications},
		title =         {Diffusion of macromolecules in the brain:
			implications for drug delivery},
		volume =        {10},
		year =          {2013},
	}

@article{RingstadEtAl2017,
		author =        {Ringstad, Geir and Vatnehol, Svein Are Sirirud and
			Eide, Per Kristian},
		journal =       {Brain},
		number =        {10},
		pages =         {2691--2705},
		publisher =     {Oxford University Press},
		title =         {{Glymphatic MRI in idiopathic normal pressure
				hydrocephalus}},
		volume =        {140},
		year =          {2017},
	}

@article{RingstadEtAl2018,
		author =        {Ringstad, Geir and Valnes, Lars M and Dale, Anders M and
			Pripp, Are H and Vatnehol, Svein-Are S and
			Emblem, Kyrre E and Mardal, Kent-Andre and
			Eide, Per K},
		journal =       {JCI Insight},
		number =        {13},
		pages =         {16},
		publisher =     {Am Soc Clin Investig},
		title =         {{Brain-wide glymphatic enhancement and clearance in
				humans assessed with MRI}},
		volume =        {3},
		year =          {2018},
	}

@article{AsgariEtAl2016,
		author =        {Asgari, Mahdi and {De Z{\'{e}}licourt}, Diane and
			Kurtcuoglu, Vartan},
		journal =       {Scientific Reports},
		pages =         {38635},
		publisher =     {Nature Publishing Group},
		title =         {{Glymphatic solute transport does not require bulk
				flow}},
		volume =        {6},
		year =          {2016},
	}

@article{sharp2019dispersion,
		author =        {Sharp, M Keith and Carare, Roxana O and
			Martin, Bryn A},
		journal =       {Fluids and Barriers of the CNS},
		number =        {1},
		pages =         {13},
		publisher =     {BioMed Central},
		title =         {{Dispersion in porous media in oscillatory flow
				between flat plates: applications to intrathecal,
				periarterial and paraarterial solute transport in the
				central nervous system}},
		volume =        {16},
		year =          {2019},
	}

@article{daversin2020mechanisms,
		author =        {Daversin-Catty, Cecile and Vinje, Vegard and
			Mardal, Kent-Andre and Rognes, Marie E},
		journal =       {bioRxiv},
		publisher =     {Cold Spring Harbor Laboratory},
		title =         {The mechanisms behind perivascular fluid flow},
		year =          {2020},
	}

@article{mestre2018flow,
		author =        {Mestre, Humberto and Tithof, Jeffrey and Du, Ting and
			Song, Wei and Peng, Weiguo and Sweeney, Amanda M and
			Olveda, Genaro and Thomas, John H and
			Nedergaard, Maiken and Kelley, Douglas H},
		journal =       {Nature Communications},
		number =        {1},
		pages =         {4878},
		publisher =     {Nature Publishing Group},
		title =         {{Flow of cerebrospinal fluid is driven by arterial
				pulsations and is reduced in hypertension}},
		volume =        {9},
		year =          {2018},
	}

@article{BedussiEtAl2017,
		author =        {Bedussi, Beatrice and Almasian, Mitra and
			de Vos, Judith and VanBavel, Ed and
			Bakker, Erik N T P},
		journal =       {Journal of Cerebral Blood Flow \& Metabolism},
		number =        {4},
		pages =         {719--726},
		publisher =     {SAGE Publications Sage UK: London, England},
		title =         {{Paravascular spaces at the brain surface: low
				resistance pathways for cerebrospinal fluid flow}},
		volume =        {38},
		year =          {2017},
	}

@article{HolterEtAl2017,
		author =        {Holter, Karl Erik and Kehlet, Benjamin and
			Devor, Anna and Sejnowski, Terrence J and
			Dale, Anders M and Omholt, Stig W and
			Ottersen, Ole Petter and Nagelhus, Erlend Arnulf and
			Mardal, Kent-Andr{\'{e}} and Pettersen, Klas H},
		journal =       {Proceedings of the National Academy of Sciences},
		number =        {37},
		pages =         {9894--9899},
		publisher =     {National Acad Sciences},
		title =         {{Interstitial solute transport in 3D reconstructed
				neuropil occurs by diffusion rather than bulk flow}},
		volume =        {114},
		year =          {2017},
	}

@article{CrociVinjeRognes2019,
		author =        {Croci, Matteo and Vinje, Vegard and Rognes, Marie E.},
		journal =       {Fluids and Barriers of the CNS},
		number =        {1},
		pages =         {32},
		publisher =     {Cold Spring Harbor Laboratory},
		title =         {{Uncertainty quantification of parenchymal tracer
				distribution using random diffusion and convective
				velocity fields}},
		volume =        {16},
		year =          {2019},
	}

@article{smith2007interstitial,
		author =        {Smith, Joshua H and Humphrey, Joseph AC},
		journal =       {Microvascular research},
		number =        {1},
		pages =         {58--73},
		publisher =     {Elsevier},
		title =         {Interstitial transport and transvascular fluid
			exchange during infusion into brain and tumor tissue},
		volume =        {73},
		year =          {2007},
	}

@article{fultz2019coupled,
		author =        {Fultz, Nina E and Bonmassar, Giorgio and
			Setsompop, Kawin and Stickgold, Robert A and
			Rosen, Bruce R and Polimeni, Jonathan R and
			Lewis, Laura D},
		journal =       {Science},
		number =        {6465},
		pages =         {628--631},
		publisher =     {American Association for the Advancement of Science},
		title =         {Coupled electrophysiological, hemodynamic, and
			cerebrospinal fluid oscillations in human sleep},
		volume =        {366},
		year =          {2019},
	}

@article{Bilston2003,
		author =        {Bilston, Lynne E. and Fletcher, David F. and
			Brodbelt, Andrew R. and Stoodley, Marcus A.},
		journal =       {Computer Methods in Biomechanics and Biomedical
			Engineering},
		number =        {4},
		pages =         {235--241},
		publisher =     {Taylor & Francis Group},
		title =         {{Arterial pulsation-driven cerebrospinal fluid flow
				in the perivascular space: a computational model}},
		volume =        {6},
		year =          {2003},
	}

@article{asgari2015astrocyte,
		author =        {Asgari, Mahdi and De Z{\'e}licourt, Diane and
			Kurtcuoglu, Vartan},
		journal =       {Scientific reports},
		pages =         {15024},
		publisher =     {Nature Publishing Group},
		title =         {How astrocyte networks may contribute to cerebral
			metabolite clearance},
		volume =        {5},
		year =          {2015},
	}

@article{BiehlerEtAl2015,
		author =        {Biehler, J. and Gee, M. W. and Wall, W. A.},
		journal =       {Biomechanics and Modelling in Mechanobiology},
		pages =         {489--513},
		title =         {{Towards efficient uncertainty quantification in
				complex and large-scale biomechanical problems based
				on a Bayesian multi-fidelity scheme}},
		volume =        {14},
		year =          {2015},
	}

@article{QuaglinoEtAl2018,
		author =        {Quaglino, A. and Pezzuto, S. and
			Koutsourelakis, P. S. and Auricchio, A. and
			Krause, R.},
		journal =       {International Journal for Numerical Methods in
			Biomedical Engineering},
		month =         {jul},
		number =        {7},
		pages =         {e2985},
		publisher =     {Wiley-Blackwell},
		title =         {{Fast uncertainty quantification of activation
				sequences in patient-specific cardiac
				electrophysiology meeting clinical time constraints}},
		volume =        {34},
		year =          {2018},
		doi =           {10.1002/cnm.2985},
		issn =          {20407939},
		url =           {http://doi.wiley.com/10.1002/cnm.2985},
	}

@misc{Quaglino2018,
		author =        {Quaglino, A. and Pezzuto, S. and Krause, R.},
		booktitle =     {Preprint},
		title =         {{Generalized multifidelity Monte Carlo estimators}},
		year =          {2018},
	}

@article{CaflishMorokoffOwen,
		author =        {Caflish, R. E. and Morokoff, W. and Owen, A.},
		journal =       {Journal of Computational Finance},
		pages =         {27--46},
		title =         {{Valuation of mortgage-backed securities using
				Brownian bridges to reduce effective dimension}},
		volume =        {1},
		year =          {1997},
	}

@article{giles2015multilevel,
		author =        {Giles, Michael B.},
		journal =       {Acta Numerica},
		pages =         {259--328},
		publisher =     {Cambridge University Press},
		title =         {{Multilevel {Monte Carlo} methods}},
		volume =        {24},
		year =          {2015},
		isbn =          {0302-9743},
	}

@article{peherstorfer2018survey,
		author =        {Peherstorfer, Benjamin and Willcox, Karen and
			Gunzburger, Max},
		journal =       {SIAM Review},
		number =        {3},
		pages =         {550--591},
		publisher =     {SIAM},
		title =         {Survey of multifidelity methods in uncertainty
			propagation, inference, and optimization},
		volume =        {60},
		year =          {2018},
	}

@article{Croci2018,
		author =        {Croci, M. and Giles, M. B. and Rognes, M. E. and
			Farrell, P. E.},
		journal =       {SIAM/ASA Journal on Uncertainty Quantification},
		number =        {4},
		pages =         {1630--1655},
		publisher =     {SIAM/ASA},
		title =         {{Efficient white noise sampling and coupling for
				multilevel Monte Carlo with non-nested meshes}},
		volume =        {6},
		year =          {2018},
	}

@article{CrociMLQMC,
		author =        {Croci, M. and Giles, M. B. and Farrell, P. E.},
		journal =       {arXiv preprint arXiv:1911.12099},
		title =         {Multilevel quasi Monte Carlo methods for elliptic
			PDEs with random field coefficients via fast white
			noise sampling},
		year =          {2019},
	}

@article{AlnaesBlechta2015a,
		author =        {Aln{\ae}s, Martin S and Blechta, Jan and Hake, Johan and
			Johansson, August and Kehlet, Benjamin and
			Logg, Anders and Richardson, Chris and Ring, Johannes and
			Rognes, Marie E and Wells, Garth N},
		journal =       {Archive of Numerical Software},
		number =        {100},
		title =         {{The FEniCS Project Version 1.5}},
		volume =        {3},
		year =          {2015},
	}

@misc{libsupermesh,
		author =        {Maddison, James R. and Panourgias, Iakovos and
			Farrell, Patrick Emmet},
		title =         {{Libsupermesh}},
		year =          {2017},
		url =           {https://bitbucket.org/libsupermesh/libsupermesh},
	}

@techreport{libsupermesh-tech-report,
		author =        {Maddison, James R. and Farrell, P. E. and
			Panourgias, Iakovos},
		institution =   {Archer, eCSE03-08},
		title =         {{Parallel supermeshing for multimesh modelling}},
		year =          {2013},
	}

@misc{femlmc,
		author =        {Croci, M.},
		title =         {{FEMLMC - A library of finite element (quasi) Monte
				Carlo methods for FEniCS}},
		year =          {2019},
		url =           {https://bitbucket.org/croci/femlmc/},
	}

@book{Lemieux2009,
		author =        {Lemieux, C.},
		edition =       {1},
		publisher =     {Springer-Verlag New York},
		series =        {Springer Series in Statistics},
		title =         {{Monte Carlo and Quasi-Monte Carlo Sampling}},
		year =          {2009},
		isbn =          {978-0-387-78164-8},
	}

@article{Morokoff1995,
		author =        {Morokoff, William J. and Caflisch, Russel E.},
		journal =       {Journal of Computational Physics},
		number =        {2},
		pages =         {218--230},
		publisher =     {Academic Press},
		title =         {{Quasi-Monte Carlo integration}},
		volume =        {122},
		year =          {1995},
	}

@article{Owen2003,
		author =        {Owen, Art B.},
		journal =       {ACM Transactions on Modeling and Computer Simulation},
		number =        {4},
		pages =         {363--378},
		title =         {{Variance and discrepancy with alternative
				scramblings}},
		volume =        {13},
		year =          {2003},
	}

@incollection{Heinrich2001,
		address =       {Berlin, Heidelberg},
		author =        {Heinrich, Stefan},
		booktitle =     {Large-Scale Scientific Computing},
		pages =         {58--67},
		publisher =     {Springer Berlin Heidelberg},
		title =         {{Multilevel Monte Carlo Methods}},
		year =          {2001},
	}

@article{giles2008,
		author =        {Giles, Michael B},
		journal =       {Operations Research},
		number =        {3},
		pages =         {607--617},
		title =         {{Multilevel Monte Carlo path simulation}},
		volume =        {56},
		year =          {2008},
	}

@article{haji2016optimization,
		author =        {Haji-Ali, Abdul-Lateef and Nobile, Fabio and
			von Schwerin, Erik and Tempone, Ra{\'{u}}l},
		journal =       {Stochastics and Partial Differential Equations
			Analysis and Computations},
		number =        {1},
		pages =         {76--112},
		publisher =     {Springer},
		title =         {{Optimization of mesh hierarchies in multilevel Monte
				Carlo samplers}},
		volume =        {4},
		year =          {2016},
	}

@article{FeischlKuoSloan2018,
		author =        {Feischl, Michael and Kuo, Frances Y and Sloan, Ian H},
		journal =       {Numerische Mathematik},
		pages =         {639--676},
		title =         {{Fast random field generation with H-matrices}},
		volume =        {140},
		year =          {2018},
	}

@article{Khoromskij2009,
		author =        {Khoromskij, B N and Litvinenko, A and Matthies, H G and
			Litvinenko, A and Matthies, H G},
		journal =       {Computing},
		number =        {1-2},
		pages =         {49--67},
		title =         {{Application of hierarchical matrices for computing
				the Karhunen-Lo{\`{e}}ve expansion}},
		volume =        {84},
		year =          {2009},
	}

@book{Hackbush2015HMatrices,
		author =        {Hackbusch, Wolfgang},
		pages =         {511},
		publisher =     {Springer},
		title =         {{Hierarchical Matrices: Algorithms and Analysis}},
		year =          {2015},
		isbn =          {978-3-662-47323-8},
	}

@article{DolzHarbrechtSchwab2017,
		author =        {D\"olz, J. and Harbrecht, H. and Schwab, Ch.},
		journal =       {Numerische Mathematik},
		number =        {4},
		pages =         {1045--1071},
		publisher =     {Springer Berlin Heidelberg},
		title =         {{Covariance regularity and H-matrix approximation for
				rough random fields}},
		volume =        {135},
		year =          {2017},
	}

@book{sullivan2015introduction,
		author =        {Sullivan, Timothy John},
		publisher =     {Springer},
		title =         {{Introduction to Uncertainty Quantification}},
		volume =        {63},
		year =          {2015},
		isbn =          {978-3-319-23394-9},
	}

@article{rodriguez2019uncertainty,
		author =        {Rodr{\'\i}guez-Cantano, Roc{\'\i}o and
			Sundnes, Joakim and Rognes, Marie E},
		journal =       {International journal for numerical methods in
			biomedical engineering},
		number =        {5},
		pages =         {e3178},
		publisher =     {Wiley Online Library},
		title =         {Uncertainty in cardiac myofiber orientation and
			stiffnesses dominate the variability of left
			ventricle deformation response},
		volume =        {35},
		year =          {2019},
	}

@article{WoodChan1994,
		author =        {Wood, A. T. A. and Chan, G.},
		journal =       {Journal of Computational and Graphical Statistics},
		number =        {4},
		pages =         {409--432},
		title =         {{Simulation of stationary Gaussian processes in
				$[0,1]^d$}},
		volume =        {3},
		year =          {1994},
	}

@article{dietrich1997fast,
		author =        {Dietrich, C R and Newsam, Garry Neil},
		journal =       {SIAM Journal on Scientific Computing},
		number =        {4},
		pages =         {1088--1107},
		publisher =     {SIAM},
		title =         {{Fast and exact simulation of stationary Gaussian
				processes through circulant embedding of the
				covariance matrix}},
		volume =        {18},
		year =          {1997},
	}

@article{GrahamKuoEtAl2018,
		author =        {Graham, I G and Kuo, F Y and Nuyens, D and
			Scheichl, R and Sloan, I H},
		journal =       {SIAM Journal on Numerical Analysis},
		number =        {3},
		pages =         {1871--1895},
		publisher =     {SIAM},
		title =         {{Analysis of circulant embedding methods for sampling
				stationary random fields}},
		volume =        {56},
		year =          {2018},
	}

@misc{BachmayrGraham2019,
		author =        {Bachmayr, Markus and Graham, Ivan G. and
			Nguyen, Van Kien and Scheichl, Robert},
		booktitle =     {Preprint},
		title =         {{Unified Analysis of Periodization-Based Sampling
				Methods for Mat\'ern Covariances}},
		year =          {2019},
	}

@article{Whittle1954,
		author =        {{P. Whittle}},
		journal =       {Biometrika},
		number =        {3-4},
		pages =         {434--449},
		title =         {{On stationary processes in the plane}},
		volume =        {41},
		year =          {1954},
	}

@article{Lindgren2011,
		author =        {Lindgren, Finn and Rue, H{\aa}vard and
			Lindstr{\"{o}}m, Johan},
		journal =       {Journal of the Royal Statistical Society: Series B
			(Statistical Methodology)},
		number =        {4},
		pages =         {423--498},
		publisher =     {Wiley for the Royal Statistical Society},
		title =         {{An explicit link between Gaussian fields and
				Gaussian Markov random fields: the stochastic partial
				differential equation approach}},
		volume =        {73},
		year =          {2009},
	}

@misc{Khristenko2018,
		author =        {Khristenko, U and Scarabosio, L and Swierczynski, P and
			Ullmann, E and Wohlmuth, B},
		booktitle =     {Preprint},
		title =         {{Analysis of boundary effects on PDE-based sampling
				of Whittle-Mat{\'{e}}rn random fields}},
		year =          {2018},
	}

@article{FangEtAl2010,
		author =        {Fang, Qianqian},
		journal =       {Biomedical Optics Express},
		number =        {1},
		pages =         {165--175},
		publisher =     {Optical Society of America},
		title =         {{Mesh-based Monte Carlo method using fast ray-tracing
				in Pl{\"{u}}cker coordinates}},
		volume =        {1},
		year =          {2010},
	}

@article{Nicholson2001,
		author =        {Nicholson, Charles},
		journal =       {Reports on Progress in Physics},
		number =        {7},
		pages =         {815},
		publisher =     {IOP Publishing},
		title =         {{Diffusion and related transport mechanisms in brain
				tissue}},
		volume =        {64},
		year =          {2001},
	}

@article{orevskovic2017new,
		author =        {Ore{\v{s}}kovi{\'{c}}, Darko and Rado{\v{s}}, Milan and
			Klarica, Marijan},
		journal =       {Pediatric Neurosurgery},
		number =        {6},
		pages =         {417--425},
		publisher =     {Karger Publishers},
		title =         {{New concepts of cerebrospinal fluid physiology and
				development of hydrocephalus}},
		volume =        {52},
		year =          {2017},
	}

@book{Wood2013,
		author =        {Wood, James H},
		edition =       {2},
		publisher =     {Springer Science \& Business Media},
		title =         {{Neurobiology of Cerebrospinal Fluid}},
		year =          {2013},
		isbn =          {978-1-4615-9271-6},
	}

@article{tuch2002high,
		author =        {Tuch, David S and Reese, Timothy G and
			Wiegell, Mette R and Makris, Nikos and
			Belliveau, John W and Wedeen, Van J},
		journal =       {Magnetic Resonance in Medicine},
		number =        {4},
		pages =         {577--582},
		publisher =     {Wiley Online Library},
		title =         {{High angular resolution diffusion imaging reveals
				intravoxel white matter fiber heterogeneity}},
		volume =        {48},
		year =          {2002},
	}

@article{KiviniemiEtAl2016,
		author =        {Kiviniemi, Vesa and Wang, Xindi and Korhonen, Vesa and
			Kein{\"{a}}nen, Tuija and Tuovinen, Timo and
			Autio, Joonas and LeVan, Pierre and Keilholz, Shella and
			Zang, Yu-Feng and Hennig, J{\"{u}}rgen and Others},
		journal =       {Journal of Cerebral Blood Flow \& Metabolism},
		number =        {6},
		pages =         {1033--1045},
		publisher =     {SAGE Publications Sage UK: London, England},
		title =         {{Ultra-fast magnetic resonance encephalography of
				physiological brain activity--Glymphatic pulsation
				mechanisms?}},
		volume =        {36},
		year =          {2016},
	}

@article{rajna20193d,
		author =        {Rajna, Zal{\'{a}}n and Raitamaa, Lauri and
			Tuovinen, Timo and Heikkil{\"{a}}, Janne and
			Kiviniemi, Vesa and Sepp{\"{a}}nen, Tapio},
		journal =       {IEEE Transactions on Medical Imaging},
		number =        {9},
		pages =         {9},
		publisher =     {IEEE},
		title =         {{3D multi-resolution optical flow analysis of
				cardiovascular pulse propagation in human brain}},
		volume =        {38},
		year =          {2019},
	}

@book{nelsen2007introduction,
		author =        {Nelsen, Roger B},
		publisher =     {Springer Science \& Business Media},
		title =         {{An Introduction to Copulas}},
		year =          {2007},
		isbn =          {978-0-387-28659-4},
	}

@book{PetterAbrahamsen1997,
		author =        {Abrahamsen, Petter},
		edition =       {2},
		publisher =     {Norwegian Computing Center},
		title =         {{A Review of Gaussian Random Fields and Correlation
				Functions}},
		year =          {1997},
		isbn =          {8-25-390435-5},
	}

@article{albargothy2018convective,
		author =        {Albargothy, Nazira J and Johnston, David A and
			MacGregor-Sharp, Matthew and Weller, Roy O and
			Verma, Ajay and Hawkes, Cheryl A and
			Carare, Roxana O},
		journal =       {Acta Neuropathologica},
		number =        {1},
		pages =         {139--152},
		publisher =     {Springer},
		title =         {{Convective influx/glymphatic system: tracers
				injected into the CSF enter and leave the brain along
				separate periarterial basement membrane pathways}},
		volume =        {136},
		year =          {2018},
	}

@book{Evans2010,
		author =        {Evans, Lawrence C.},
		publisher =     {American Mathematical Society},
		title =         {{Partial Differential Equations}},
		year =          {2010},
		isbn =          {0-8218-4974-3},
	}

@article{gmsh,
		author =        {Geuzaine, Christophe and
			Remacle, Jean-Fran{\c{c}}ois},
		journal =       {International Journal for Numerical Methods in
			Engineering},
		number =        {11},
		pages =         {1309--1331},
		publisher =     {Wiley Online Library},
		title =         {{Gmsh: A 3-D finite element mesh generator with
				built-in pre-and post-processing facilities}},
		volume =        {79},
		year =          {2009},
	}

@inproceedings{thomee2014positivity,
		author =        {Thom{\'{e}}e, Vidar},
		booktitle =     {International Conference on Numerical Methods and
			Applications},
		organization =  {Springer},
		pages =         {13--24},
		title =         {{On positivity preservation in some finite element
				methods for the heat equation}},
		year =          {2014},
	}

@book{LoggEtAl2012,
		author =        {Logg, Anders and Mardal, Kent-Andre and Wells, Garth},
		publisher =     {Springer Science \& Business Media},
		title =         {{Automated Solution of Differential Equations by the
				Finite Element Method: the FEniCS Book}},
		volume =        {84},
		year =          {2012},
		isbn =          {978-3-642-23098-1},
	}

@techreport{balay2014petsc,
		author =        {Balay, S. and Abhyankar, S. and Adams, M. and
			Brown, J. and Brune, P R and Buschelman, K and
			Eijkhout, V and Gropp, W and Kaushik, D and
			Knepley, M G and Others},
		institution =   {Argonne National Laboratory (ANL)},
		title =         {{{PET}Sc users manual revision 3.8}},
		year =          {2017},
	}

@inproceedings{hypre,
		author =        {Falgout, Robert D. and Yang, Ulrike Meier},
		booktitle =     {International Conference on Computational Science},
		pages =         {632--641},
		publisher =     {Springer},
		title =         {{\textit{hypre}: A library of high performance
				preconditioners}},
		year =          {2002},
	}

@article{Cliffe2011,
		author =        {Cliffe, K. A. and Giles, M. B. and Scheichl, R. and
			Teckentrup, A. L.},
		journal =       {Computing and Visualization in Science},
		number =        {1},
		pages =         {3--15},
		publisher =     {Springer},
		title =         {{Multilevel Monte Carlo methods and applications to
				elliptic PDEs with random coefficients}},
		volume =        {14},
		year =          {2011},
	}

@article{TeckentrupMLMC2013,
		author =        {Teckentrup, A. L. and Scheichl, R. and Giles, M. B. and
			Ullmann, E.},
		journal =       {Numerische Mathematik},
		number =        {3},
		pages =         {569--600},
		title =         {{Further analysis of multilevel Monte Carlo methods
				for elliptic PDEs with random coefficients}},
		volume =        {125},
		year =          {2013},
	}

@book{brenner2007mathematical,
		author =        {Brenner, Susanne and Scott, Ridgway L.},
		publisher =     {Springer Science and Business Media},
		title =         {{The Mathematical Theory of Finite Element Methods}},
		volume =        {15},
		year =          {1994},
		isbn =          {978-1-4757-4340-1},
	}

@incollection{GilesWaterhouse2009,
		author =        {Giles, M. B. and Waterhouse, B. J.},
		booktitle =     {Advanced Financial Modelling (Radon Series on
			Computational and Applied Mathematics)},
		number =        {8},
		pages =         {165--181},
		publisher =     {De Gruyter},
		title =         {{Multilevel quasi-Monte Carlo path simulation}},
		year =          {2009},
		isbn =          {978-3110213133},
	}

@article{lemaire2017multilevel,
		author =        {Lemaire, Vincent and Pag{\`{e}}s, Gilles},
		journal =       {Bernoulli},
		number =        {4A},
		pages =         {2643--2692},
		publisher =     {Bernoulli Society for Mathematical Statistics and
			Probability},
		title =         {{Multilevel Richardson-Romberg extrapolation}},
		volume =        {23},
		year =          {2017},
	}

@article{CharrierMLMC2013,
		author =        {Charrier, Julia and Scheichl, Robert and
			Teckentrup, Aretha L},
		journal =       {SIAM Journal of Numerical Analysis},
		number =        {1},
		pages =         {322--352},
		title =         {{Finite element error analysis of elliptic PDEs with
				random coefficients and its application to multilevel
				Monte Carlo methods}},
		volume =        {51},
		year =          {2013},
	}

@article{KuoSchwabSloan2015,
		author =        {Kuo, F. Y. and Schwab, C. and Sloan, I. H.},
		journal =       {Foundations of Computational Mathematics},
		number =        {2},
		pages =         {411--449},
		title =         {{Multi-level quasi-Monte Carlo finite element methods
				for a class of elliptic PDEs with random
				coefficients}},
		volume =        {15},
		year =          {2015},
	}

@article{KuoScheichlSchwabEtAl2017,
		author =        {Kuo, F. Y. and Scheichl, R. and Schwab, Ch. and
			Sloan, I. H. and Ullmann, E.},
		journal =       {Mathematics of Computation},
		number =        {308},
		pages =         {2827--2860},
		title =         {{Multilevel quasi-Monte Carlo methods for lognormal
				diffusion problems}},
		volume =        {86},
		year =          {2017},
	}

@misc{colin27,
		author =        {Fang, Qianqian},
		title =         {{Colin27 adult brain atlas FEM mesh}},
		year =          {2011},
		url =           {http://mcx.sourceforge.net/},
	}

@inproceedings{potsepaev2010application,
		author =        {Potsepaev, R. and Farmer, Chris L.},
		booktitle =     {ECMOR XII-12th European Conference on the Mathematics
			of Oil Recovery},
		number =        {September},
		title =         {{Application of stochastic partial differential
				equations to reservoir property modelling}},
		volume =        {2},
		year =          {2014},
		isbn =          {978-1-6343-9168-9},
	}

\end{document}